\documentclass[twocolumn,amssymb,amsmath,pra, floatfix]{revtex4-1}

\usepackage{physics}
\usepackage{graphicx} 
\usepackage[caption=false, labelformat=simple, position=top]{subfig}
\usepackage{dcolumn}

\usepackage[colorlinks=true,linkcolor = {blue}, citecolor = {blue}, urlcolor = {blue}]{hyperref}
\usepackage[dvipsnames]{xcolor}

\newcommand{\kristian}[1]{{\color{blue}#1}}

\begin{document}

\title{Holes and magnetic polarons  in a triangular lattice antiferromagnet}
\author{Jasper van de Kraats}
\affiliation{Eindhoven University of Technology, P. O. Box 513, 5600 MB Eindhoven, The Netherlands}
\author{Kristian K.\ Nielsen}
\affiliation{Max-Planck Institute for Quantum Optics, Hans-Kopfermann-Str. 1, D-85748 Garching, Germany}
\author{Georg M.\ Bruun}
\email[]{bruungmb@phys.au.dk}
\affiliation{Center for Complex Quantum Systems, Department of Physics and Astronomy, Aarhus University, Ny Munkegade, DK-8000 Aarhus C, Denmark. }
\affiliation{Shenzhen Institute for Quantum Science and Engineering and Department of Physics, Southern University of Science and Technology, Shenzhen 518055, China}

\begin{abstract}
The intricate interplay between charge motion and magnetic order in geometrically frustrated lattices is central for the properties of many two-dimensional quantum materials. The triangular lattice antiferromagnet is a canonical example of a frustrated system, and here we analyse the dynamics of a hole in such a lattice focusing on observables that have become accessible in a new generation of  experiments. Using the $t$-$J$ model, we solve the problem exactly within linear spin wave theory in the limit of strong magnetic interactions, showing that the ground state is described by a coherent state of spin waves. The derivation highlights the crucial role played by the interaction  between a static hole and the neighboring spins, which originates in the geometric frustration and has often been omitted in earlier works. Furthermore, we show that the non-equilibrium dynamics after a  hole has abruptly been inserted at a  lattice site is given exactly by a coherent state with time-dependent oscillatory coefficients. 
Physically, this describes a burst of magnetic frustration propagating through only two-thirds of the lattice sites, since a destructive interference of spin waves leaves spins parallel to that removed by the hole  unperturbed. After the wave has propagated through the lattice, the magnetization relaxes to that of the ground state. We then use our analytical solution to benchmark the widely used self-consistent Born approximation (SCBA), showing that it is very accurate also for a triangular lattice. The magnetic polaron spectrum is analysed for general magnetic interactions using the SCBA, and we compare our results with those for a square lattice. 
  \end{abstract}
  
  \maketitle

\section{Introduction}
\label{sec:intro}
Understanding the competition between hole motion and magnetic order  is a major challenge.  
The small doping limit of high-temperature superconductors is characterised by holes moving in a square lattice antiferromagnet (SAFM), such that a 
description of these processes constitute an important step towards understanding these complicated materials~\cite{Anderson1987, Lee2006,Emery1987, Schrieffer1988, Dagotto1994}. 
Partly due to this connection,  the vast majority of  studies have focused on  hole dynamics in a SAFM. The properties of a hole in other lattices 
  is however also of fundamental interest. It was suggested 
that the inherent geometrical  frustration of a two-dimensional 
triangular lattice leads to the formation of a resonating valence bond state with no long range order~\cite{Anderson1973}.
While such quantum liquid states may be realised  for intermediate coupling strengths~\cite{Szasz2020}, 
it is now widely recognised from series expansions as well as numerical 
calculations that the ground state of a triangular lattice
for strong coupling has long-range antiferromagnetic order based on the $120^{\circ}$ antiferromagnetic 
N\'eel state~\cite{Capriotti1999,Zheng2006,White2007}. The increased role of fluctuations due to frustration 
has been shown to lead to interesting effects on the spin wave spectrum of such a triangular lattice antiferromagnet 
(TAFM)~\cite{Leung1993, Chernyshev2009}. It also gives rise to perculiar properties of hole dynamics and magnetic polarons in 
TAFMs~\cite{Azzouz1996, Apel1998, Vojta1999, Srivastava2005,Trumper2004}. 
Several crystals realise a triangular lattice, but a  microscopic description of their properties remains an open question due to their complicated 
nature with many unknown 
parameters~\cite{Shimizu2003,Maska2004, Wang2004,Yamashita:2008uz,Itou2008,Law2017,Li2018,Ni2019,Bourgeois2019,Zhou2017}.

Recent experiments using cold atoms in optical lattices have provided  a wealth of  new  information regarding 
 the motion of holes in fermionic spin systems~\cite{Christie2019,Brown2019,Koepsell:2019ua,Ji2021,Koepsell2021}. These experiments 
realise the Fermi-Hubbard model essentially perfectly and, moreover, their  single site resolution gives access to 
the  real space dynamics of fermions in the lattice. A new generation of experiments trapping bosonic~\cite{Yamamoto_2020} and  fermionic~\cite{Yang2021} atoms in
 triangular optical lattices  promise to provide new and detailed experimental insights into magnetic frustration and hole dynamics.
 Moreover, recent breakthrough experiments using multilayers of atomically thin van der Waals materials have realised triangular moir\'e superlattices with tuneable Hubbard
 parameters, thereby opening up an exciting new platform for exploring geometrically frustrated lattices~\cite{Cao:2018wy,Tang:2020uf,Balents:2020wp,Wu2018}.

Inspired by these developments, in this work we explore the properties of a hole in a TAFM as described by the $t$-$J$ model. 
We show that in the limit of strong magnetic interactions and within linear spin wave theory, the exact ground state is a coherent state describing a static hole dressed  by spin waves. The dressing is due to an interaction  between the  hole and its neighbouring  spins coming from the geometric frustration of the lattice, and it has no analogy for a bi-partite lattice. We then extend our exact results to the non-equilibrium dynamics following a hole injected into a lattice site, and show that the  resulting propagation of spin waves through the lattice  is described by a time-dependent coherent state. Interestingly, spins parallel to the spin removed by the hole are unaffected due to destructive interference, so that the wave of magnetic disorder only propagates through two-thirds of the lattice. Eventually, the magnetic disorder relaxes to that of the ground state. Our analytical solution enables us to benchmark the SCBA, which is known to be accurate on a square lattice, and we show that this holds for a triangular lattice as well. Finally, we analyse the properties of a hole and the formation of magnetic polarons for general interaction strengths and compare to the case of a square lattice.  
  
The paper is structured as follows. We formulate the model in Sec.~\ref{sec:ham} and    apply the slave-fermion representation together with 
linear spin-wave theory. In Sec. \ref{sec:lJ}, we derive analytical solutions for the ground state and the non-equilibrium dynamics following a hole suddenly 
created at a given lattice site. The SCBA is introduced in  Sec.~\ref{sec:SCBA}, where we compare it to the exact solution and  use it to analyse magnetic 
polarons for general interaction strengths. Finally, we conclude and provide an outlook in Sec.~\ref{sec:conc}.

\section{Hole spin-wave Hamiltonian}
\label{sec:ham}
 We consider the $t$-$J$ model with the Hamiltonian $H = H_t + H_J$, where~\cite{Chao1977, MacDonald1988, Eskes1994}
\begin{align}
\hat{H}_t = -t \sum_{\langle \vb{i, j} \rangle} \sum_{\sigma} \left[\tilde{c}_{\vb{i, \sigma}}^{\dagger} \tilde{c}_{\vb{j,\sigma}} + \mathrm{h.c.} \right]
\label{eq:Ht}
\end{align}
describes the electron hopping. The modified creation operator 
$\tilde{c}_{\vb{i} \sigma}^{\dagger} = \left(1 -  \hat{n}_{\vb{i},\bar{\sigma}}\right) c_{\vb{i},\sigma}^{\dagger}$ 
with $\hat{n}_{\vb{i},\sigma} = \hat{c}_{\vb{i},\sigma}^{\dagger} \hat{c}_{\vb{i},\sigma}$  the local  number operator and $\bar\sigma$ the opposite spin of $\sigma$,
is defined such that $H_t$ is restricted to the Hilbert space of singly occupied sites. Here, $\hat{c}_{\vb{i}, \sigma}^{\dagger}$ create a fermion at lattice site $\vb{i}$ and spin $\sigma$. The $t$-$J$ model, therefore, naturally describes a system in which the  repulsion between spin $1/2$ fermions is much larger than the available energy. The second term $\hat{H}_J$ quantifies an antiferromagnetic Heisenberg spin-exchange interaction, 
\begin{align}
\hat{H}_J = J \sum_{\langle \vb{i, j} \rangle} \left[\hat{S}_{\vb{i}}^z \hat{S}_{\vb{j}}^z + \frac{1}{2} \left(\hat{S}_{\vb{i}}^{+} \hat{S}_{\vb{j}}^{-} -  \hat{S}_{\vb{j}}^{+} \hat{S}_{\vb{i}}^{-} \right) - \frac{\hat{n}_{\vb{i}} \hat{n}_{\vb{j}}}{4} \right].
\label{eq:HJ}
\end{align}
Here $\hat{n}_{\vb{i}} = \hat{n}_{\vb{i},\uparrow} + \hat{n}_{\vb{i},\downarrow}$, and $J > 0$ denotes the antiferromagnetic exchange energy. The spin operators are defined in the Schwinger representation as
\begin{align}
\vb{\hat{S}}_i = \frac{1}{2} \sum_{\sigma \sigma'} \hat{c}_{i,\sigma}^{\dagger} \vb{\sigma}_{\sigma \sigma'} \hat{c}_{i,\sigma'}
\end{align}
with $\vb{\sigma} = \left(\sigma^x, \sigma^y, \sigma^z \right)$ the vector of Pauli matrices.

While the triangular lattice with antiferromagnetic coupling was originally suggested to realise a resonating valence bond ground state~\cite{Anderson1973}, 
it is now known from series expansions as well as numerical 
calculations that its ground state has long-range antiferromagnetic order based on the $120^{\circ}$ antiferromagnetic 
N\'eel state~\cite{Capriotti1999,Zheng2006,White2007}. This state may be generated in the $(x,y)$ plane from the ordering vector $\vb{Q}_{\mathrm{TAFM}} = \left(4\pi/3,0 \right)$, such that a spin at position $\vb{r}_i$ has an angle $\theta_{\vb{i}} = \vb{r}_i \cdot \vb{Q}_{\mathrm{TAFM}}$ relative to the $z$-axis as illustrated  in Fig.~\ref{fig:120deg}. The lattice may then be subdivided into three sublattices with ordered spin-directions differing by angles of $2\pi/3$. For the subsequent analysis, it is convenient to perform a local coordinate rotation by $\theta_{\vb{i}}$ so that all spins point along the same local coordinate axis. The associated SU(2) rotation matrix reads~\cite{Chernyshev2009}
\begin{align}
R_{\vb{i}} = \begin{bmatrix}
\cos(\theta_{\vb{i}}/{2}) & \sin(\theta_{\vb{i}}/{2} ) \\ 
-\sin(\theta_{\vb{i}}/{2} ) & \cos(\theta_{\vb{i}}/{2} ) 
\end{bmatrix}
\end{align}
and the rotated (primed) operators are
\begin{align}
\hat{c}_{\vb{i},\sigma}^{\dagger \prime} = \sum_{\sigma'} R_{\vb{i},\sigma \sigma'} \hat{c}_{\vb{i},\sigma'}^{\dagger}, \qquad \vb{\hat{S}}_i' = R_{\vb{i}}^{-1} \vb{\hat{S}}_{\vb{i}} R_{\vb{i}}.
\end{align}
Upon substitution into equations \eqref{eq:Ht} and \eqref{eq:HJ} one obtains a rotated $t$-$J$ model for which the Néel state is now an effective ferromagnet. This prescription is straightforwardly applied to the SAFM as well, with altered ordering vector $\vb{Q}_{\mathrm{SAFM}} = \left(\pi, \pi\right)$. We use units where the lattice constant and $\hbar$ both are unity. 

\subsection{Slave-fermion representation}
To model the dynamics of the hole we adopt a slave-fermion representation \cite{Schmitt1988,Kane1989, Martinez1991, Liu1991, Nielsen2021}, which describes the system in terms of spinless fermionic holes created by the operators $\hat{h}_{\vb{i}}^{\dagger}$,
 and bosonic spin excitations created by the operators $\hat{s}_{\vb{i}}^{\dagger}$. We have 
 $\tilde{c}_{\vb{i},\downarrow}' = \hat{h}_{\vb{i}}^{\dagger} \hat{s}_{\vb{i}}$ and $\tilde{c}_{\vb{i}, \uparrow}' = \hat{h}_{\vb{i}}^{\dagger} \sqrt{1 - \hat{h}_{\vb{i}}^{\dagger} \hat{h}_{\vb{i}} - \hat{s}_{\vb{i}}^{\dagger} \hat{s}_{\vb{i}}}$, where the square root factor ensures that $\tilde{c}_{\vb{i}, \uparrow}$ is uneffective on any site that already contains a hole or spin-excitation. The spin-operators are written as $\hat{S}_{\vb{i}}^{+\prime} = \sqrt{1 - \hat{h}_{\vb{i}}^{\dagger} \hat{h}_{\vb{i}} - \hat{s}_{\vb{i}}^{\dagger} \hat{s}_{\vb{i}}} \hat{s}_{\vb{i}}$ 
 and $\hat{S}_{\vb{i}}^{z\prime} = (1 - \hat{h}_{\vb{i}}^{\dagger} \hat{h}_{\vb{i}} - 2\hat{s}_{\vb{i}}^{\dagger} \hat{s}_{\vb{i}})/2$. 
 Substituting into the rotated $t$-$J$ model and truncating to second order we obtain 
\begin{align}
\begin{split}
\hat{H}_J = \hat{H}_{\text{s}} + \frac{J}{4} \sum_{\langle\vb{i},\vb{j} \rangle} \sin\theta_{ij} \left[\bar{s}_{\vb{i}} 
\hat{h}_{\vb{j}}^{\dagger} \hat{h}_{\vb{j}} - \hat{h}_{\vb{i}}^{\dagger}\hat{h}_{\vb{i}} \bar{s}_{\vb{j}} \right]
\end{split}
\label{eq:HJcubic}
\end{align}
with $\theta_{ij}=\theta_{\vb{i}}-\theta_{\vb{j}}$.
Here $\bar{s}_{\vb{i}} = \hat{s}_{\vb{i}} + \hat{s}_{\vb{i}}^{\dagger}$, and $\hat{H}_{\text{s}}$ is the quadratic spin-wave Hamiltonian that also appears in the usual Holstein-Primakoff transformation on the triangular lattice \cite{Chernyshev2009}. Expressions for $\hat{H}_s$ and the hopping term $\hat{H}_t$ in the slave fermion representation  are given in App. \ref{ap:Hposfull}.

\begin{figure}
\centering
\subfloat[\label{fig:120deg}]{
\includegraphics[width=0.22\textwidth]{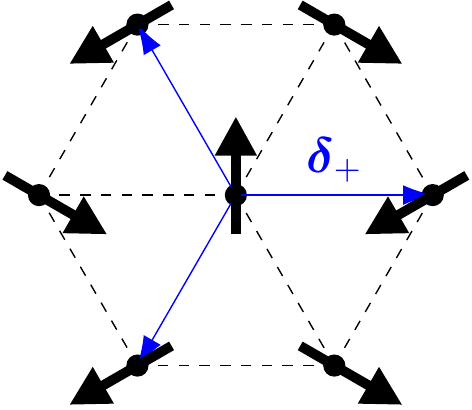}}
\subfloat[\label{fig:frustration}]{
\includegraphics[width=0.22\textwidth]{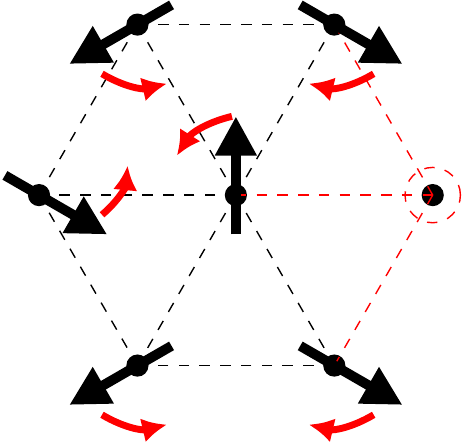}}
\caption{(a) The antiferromagnetic $120^{\circ}$ Néel state in a triangular lattice. The vectors $\vb*{\delta}_+$ that define $g_{\vb{k}}$ and $\tilde{g}_{\vb{k}}$ are shown in blue.  (b) Introducing a hole by removing a spin makes the neighbouring spins adjust their direction as indicated by the red arrows
 thereby introducing magnetic frustation. Here it is assumed that the spins exist in a much larger lattice which is not drawn.  
}
\label{fig:lattice}
\end{figure}

 The second term in Eq.~\eqref{eq:HJcubic} quantifies a $J$-dependent interaction between the holes and spin-waves, which is absent in the SAFM 
 where $\theta_{ij}  = \pm \pi$. 
 Physically, we can understand this interaction as coming from the geometric frustration on the triangular lattice.  
The ordered state of any spin is stabilised by the simultaneous spin-exchange with its six nearest neighbours. 
As we illustrate in Fig.~\ref{fig:lattice}, if one spin is removed the neighbouring  spins will adjust their direction away from the $120^\circ$  order
to minimise the energy. This corresponds to introducing spin waves in the system.  
Indeed, inspecting Eq.~\eqref{eq:HJcubic} shows that this interaction describes a spin-excitation  
arising from a  purely stationary hole in the lattice. We note that this term, which has no equivalence for a square lattice, 
 has been omitted in earlier works describing hole dynamics in TAFMs using a slave-fermion representation \cite{Azzouz1996, Trumper2004}
so that the creation of spin waves only comes from the hopping term $\hat{H}_t$, see App.~\ref{ap:Hposfull}. Such an approximation must be expected to be accurate for $J\ll t$. As we shall demonstrate below, the $J$-dependent interaction between the hole and the spin waves is however important away from this regime and significantly alters the asymptotic large $J$ behavior of the hole. 

\begin{figure*}[t]
\centering
\subfloat[\label{fig:disp_bare_0}]{
\includegraphics[width=0.45\textwidth]{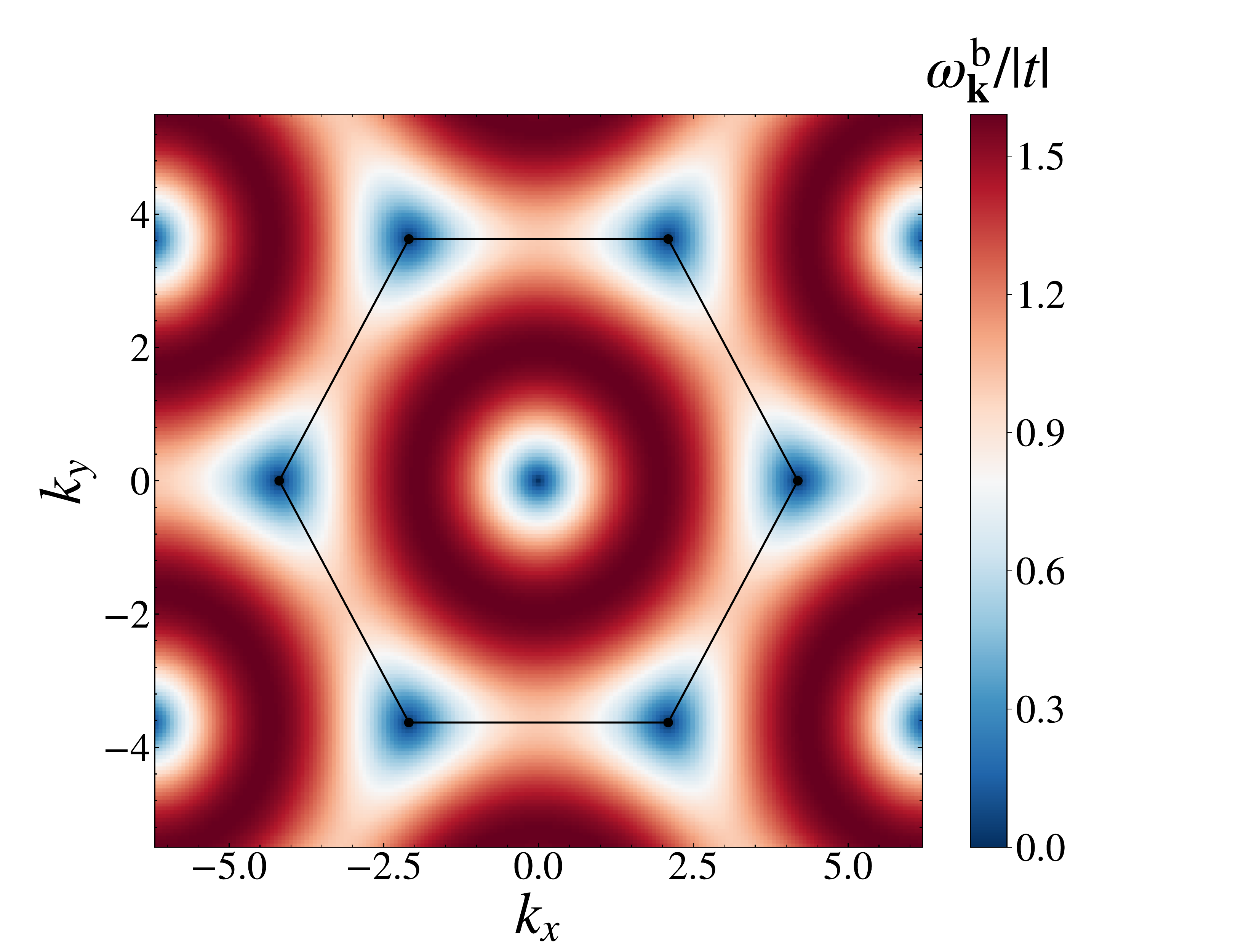}}
\subfloat[\label{fig:disp_bare_1}]{
\includegraphics[width=0.45\textwidth]{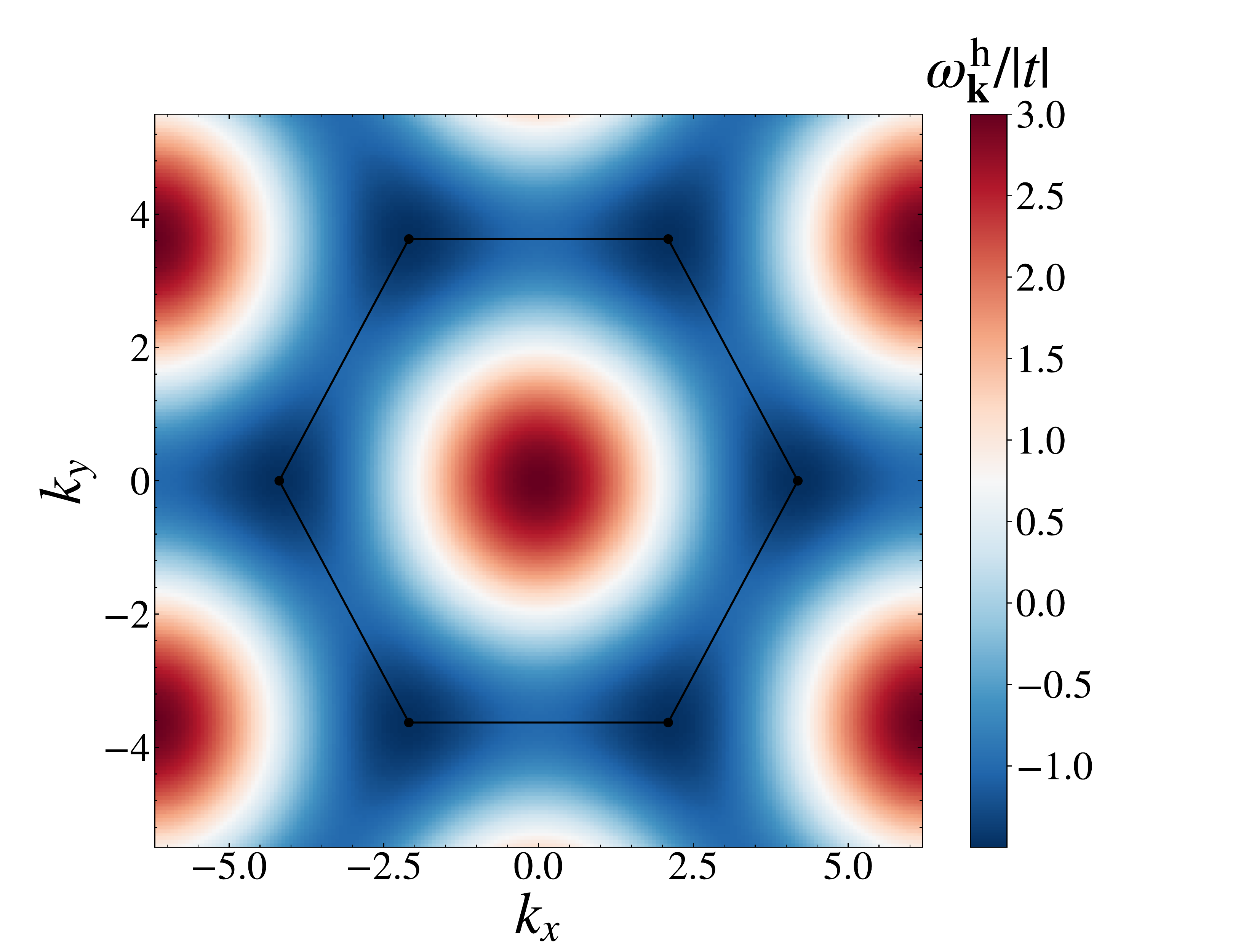}}
\caption{(a) Spin-wave dispersion $\omega_{\vb{k}}^{\mathrm{b}}$. (b) Hole dispersion $\omega_{\vb{k}}^{\mathrm{h}}$. Both for  $J/\abs{t} = 1$ and $t < 0$.
}
\label{fig:disp_bare}
\end{figure*}

Fourier transforming and diagonalizing the harmonic part of $H_J$ by the Bogoliubov transformation $\hat{s}_{\vb{k}} = u_{\vb{k}} \hat{b}_{\vb{k}} - v_{\vb{k}} \hat{b}_{-\vb{k}}^{\dagger}$, yields 
\begin{align}
\begin{split}
\hat{H} &= \sum_{\vb{k}} \omega_{\vb{k}}^{\mathrm{h}} \hat{h}_{\vb{k}}^{\dagger} \hat{h}_{\vb{k}} + \sum_{\vb{k}} \omega_{\vb{k}}^{\mathrm{b}} \hat{b}_{\vb{k}}^{\dagger} \hat{b}_{\vb{k}} \\& \qquad - \sum_{\vb{k},\vb{q}} \hat{h}_{\vb{k} + \vb{q}}^{\dagger} \hat{h}_{\vb{k}} \left[V_{\vb{k},\vb{q}} \hat{b}_{-\vb{q}}^{\dagger} - V_{\vb{k}+\vb{q}, -\vb{q}} \hat{b}_{\vb{q}} \right].
\end{split}
\label{eq:Hkspace}
\end{align}
The hole and spin-wave dispersions are 
\begin{align}
\begin{split}
\omega_{\vb{k}}^{\mathrm{h}} = -3t g_{\vb{k}}, \qquad \omega_{\vb{k}}^{\mathrm{b}} = \frac{3}{2}J \sqrt{\left(1-g_{\vb{k}} \right)\left(1 + 2g_{\vb{k}}  \right)},
\end{split}
\label{Dispersions}
\end{align}
with the structure factor $g_{\vb{k}} = 1/3 \sum_{\vb*{\delta}_+} \cos\left(\vb{k} \cdot \vb*{\delta}_{+} \right)$  where $\vb{\delta}_+$ denotes a set of three 
nearest neighbour vectors illustrated in Fig.~\ref{fig:120deg}. These dispersions are shown in Fig.~\ref{fig:disp_bare}. 
Note that sign reversal symmetry with respect to $t$ is broken in a triangular lattice as opposed to a square lattice. 
The hole dispersion term, which is absent on a SAFM, arises from the fact 
that adjacent spins on the TAFM are not anti-parallel so that a hopping hole  has a nonzero chance of projecting the displaced spin into the "correct" state that agrees with the antiferromagnetic order. Hence even if just a single hole is present it may move across the lattice without frustrating the antiferromagnetic structure, 
which is impossible on the square lattice~\cite{Trumper2004}. 

The vertex for the interaction between a hole and the spin waves is
\begin{align}
V_{\vb{k,q}}  
= &\frac{3\sqrt{3}i}{\sqrt{N}} t\left(v_{\vb{q}} \tilde{g}_{\vb{k}} + u_{\vb{q}} \tilde{g}_{\vb{k}+\vb{q}} \right) + \frac{3\sqrt{3}i}{4\sqrt{N}} J  \left(u_{\vb{q}} - v_{\vb{q}} \right)\tilde{g}_{\vb{q}}\nonumber\\
= &V_{\vb{k,q}}^t + V_{\vb{q}}^J
\label{eq:vertex}
\end{align}
with $\tilde{g}_{\vb{k}} = 1/3 \sum_{\vb*{\delta}_+} \sin\left(\vb{k} \cdot \vb*{\delta}_{+} \right)$ and 
\begin{align}
\begin{split}
u_{\vb{k}} &= \frac{1}{2} \sqrt{\frac{2+g_{\vb{k}}}{\sqrt{\left(1-g_{\vb{k}} \right)\left(1 + 2g_{\vb{k}}  \right)}} + 2}, \\ v_{\vb{k}} &= \frac{1}{2} \mathrm{sgn}(-g_{\vb{k}})  \sqrt{\frac{2+g_{\vb{k}}}{\sqrt{\left(1-g_{\vb{k}} \right)\left(1 + 2g_{\vb{k}}  \right)}} - 2}.
\end{split}
\end{align}
We see that it contains a term $V_{\vb{k,q}}^t \propto t$ arising from the hole hopping as well as a term 
$V_{\vb{q}}^J\propto J$. The latter  is absent for a SAFM and 
arises from the geometric frustration shown in Fig. \ref{fig:frustration} as discussed above.

\section{Large $J/t$ limit}
\label{sec:lJ}
In this section, we present  analytical solutions in the limit $J/t\gg 1$ both for the ground state and for a non-equilibrium situation where a hole is 
abruptly introduced in the lattice. The term $ V_{\vb{q}}^J\propto J$ in Eq.~\eqref{eq:vertex} for the interaction vertex is clearly crucial for the properties of the hole in this limit, where it dresses the hole with spin waves. When this term is omitted as done in several previous works, the large magnetic interaction suppresses the motion of the hole and the ground state simply becomes a static hole surrounded by an unperturbed AF order, as for the SAFM. We show that when it is included, the problem can be mapped onto the well known Fr\"ohlich model in the limit of infinite mass, which allows for an analytic solution describing an immobile hole surrounded by magnetic frustations. 

In the $J/t\gg 1$ limit, we can neglect $\hat H_t$  and the Hamiltonian simplifies to 
\begin{align}
\hat{H}_J = \sum_{\vb{k}} \omega_{\vb{k}}^{\mathrm{b}} \hat{b}_{\vb{k}}^{\dagger} \hat{b}_{\vb{k}}  - \sum_{\vb{k},\vb{q}} \hat{h}_{\vb{k} + \vb{q}}^{\dagger} \hat{h}_{\vb{k}} \left[V_{\vb{q}}^J \hat{b}_{\vb{q}} + V_{\vb{q}}^{J*} \hat{b}_{\vb{q}}^{\dagger}  \right].
\label{eq:HamReduced}
\end{align}
As is apparent from the real space representation of $\hat{H}_J$ given by Eq.~\eqref{eq:HJcubic}, the hole is now stationary in the lattice  where it emits or 
absorbs spin waves.  Therefore, without loss of generality, we can assume the hole to be located 
at a certain lattice site $\vb{r}$. In the Hilbert space $\hat{h}_{\vb{r}}^{\dagger} \ket{\Psi_\text{s}}$ where $\ket{\Psi_\text{s}}$ is a general spin state of the lattice surrounding the hole, 
a straightforward calculation shows that the Hamiltonian in Eq.~\eqref{eq:HamReduced} is equivalent to
\begin{align}
\hat{H}_J =  \sum_{\vb{k}}\left[ \omega_{\vb{k}}^{\mathrm{b}} \hat{b}_{\vb{k}}^{\dagger} \hat{b}_{\vb{k}}  -  V_{\vb{k}}^J \hat{b}_{\vb{k}}e^{i \vb{k}\cdot \vb{r}} + V_{\vb{k}}^{J*} \hat{b}_{\vb{k}}^{\dagger}e^{-i \vb{k}\cdot \vb{r}} \right].
\label{Frohlich}
\end{align}
After a gauge transform $\tilde{b}_{\vb{k}} = \hat{b}_{\vb{k}}e^{i \vb{k} \cdot \vb{r}}$,  we obtain  the Fr\"ohlich Hamiltonian for an infinite mass impurity~\cite{Mahan2000book}. 

\subsection{Ground state}
The Fr\"ohlich Hamiltonian for infinite mass given by Eq.~\eqref{Frohlich} can be solved analytically using the  canonical transformation
\begin{align}
\begin{split}
\hat{\mathcal{S}} = e^{\sum_{\vb{k}} \left(\alpha_{\vb{k}}^* \tilde{b}_{\vb{k}}^{\dagger} -\alpha_{\vb{k}} \tilde{b}_{\vb{k}} \right)},
\end{split}
\label{eq:S}
\end{align}
where $\alpha_{\vb{k}} = V_{\vb{k}}^J/\omega_{\vb{k}}^{\mathrm{b}}$. Indeed, 
 \begin{align}
 \hat{\mathcal{S}}^{-1}\hat{H}_J \hat{\mathcal{S}} = \sum_{\vb{k}} \omega_{\vb{k}}^{\mathrm{b}} \tilde{b}_{\vb{k}}^{\dagger} \tilde{b}_{\vb{k}} + E_g
 \end{align}
  with
\begin{align}
E_g = -\sum_{\vb{k}} \frac{\abs{V_{\vb{k}}^J}^2}{\omega_{\vb{k}}^{\mathrm{b}}}= -0.144 \ J
\label{GSenergy}
\end{align}
 the ground state energy, where the numerical value is obtained for an infinite lattice. 
 The wave function of the ground state $\ket{\Psi_{\vb{r}, g}}$ is a multimode coherent state of spin-waves,
\begin{align}
\begin{split}
\ket{\Psi_{\vb{r}, g}} =\hat{\mathcal{S}} \ket{h_{\vb{r}}}= e^{-\frac{1}{2} \sum_{\vb{k}} \abs*{\alpha_{\vb{k}}}^2} e^{\sum_{\vb{k}}\alpha_{\vb{k}}^* \tilde{b}_{\vb{k}}^{\dagger}} \ket{h_{\vb{r}}},
\end{split}
\label{eq:PolaronGroundWaveFunc}
\end{align}
with $\ket{h_{\vb{r}}}=\hat h_{\mathbf r}^\dagger\ket{{\text{AF}}}$ where $\ket{\text{AF}}$ is the antiferromagnetic ground state
defined by $\hat{b}_{\vb{k}}\ket{\mathrm{AF}}=0$. It follows from Eq.~\eqref{eq:PolaronGroundWaveFunc} that
 the ground state  is  an eigenfunction of $b_{\vb{k}}$ with eigenvalue $\alpha_{\vb{k}}^*$ and that the spin-wave distribution in any given mode is Poissonian
\begin{align}
\begin{split}
\bar{n}_{\vb{k},g} = \abs*{\alpha_{\vb{k}}}^2, \qquad \sigma_{\vb{k},g} = \sqrt{\bar{n}_{\vb{k},g}}.
\end{split}
\label{eq:Groundbnumber}
\end{align}
Here $\bar{n}_{\vb{k},g}$ denotes the mean spin-wave number in mode $\vb{k}$ and $\sigma_{\vb{k},g}$ the standard deviation. From 
Eq.~\eqref{eq:PolaronGroundWaveFunc}, it is also straightforward to compute the quasiparticle residue   
\begin{align}
Z = \abs*{\braket*{h_{\mathbf r}}{\Psi_{\vb{r},g}}}^2 = e^{-\sum_{\vb{k}} \abs*{\alpha_{\vb{k}}}^2}=0.8003,
\label{ResidueExact}
\end{align}
 which quantifes the overlap of the  ground state with the  state of a localised hole  surrounded by unperturbed  AF order. 
 Since $Z>0$, the ground state corresponds to a well-defined quasiparticle, i.e.\ a magnetic polaron with infinite mass.  
 The value  $Z = 0.8003$ should be contrasted to the case of a  SAFM, where $Z \rightarrow 1$ for $J/|t|\rightarrow\infty$.
 This is because  a stationary hole introduces magnetic frustration in a TAFM as shown in Fig.~\ref{fig:lattice}, in contrast to the case of a  SAFM,.

\subsection{Time-dependent many-body wave function}
We now show that in addition to the ground state, we can also derive an analytical solution for the non-equilibrium many-body
dynamics after a hole is abruptly introduced at a given lattice site. Such a quench experiment was  recently 
performed for a  SAFM~\cite{Ji2021}. 

We imagine a hole created at   lattice site ${\mathbf r}$ at time $\tau = 0$ so that the initial state is 
$\ket{h_{\vb{r}}}=\hat h_{\mathbf r}^\dagger\ket{{\text{AF}}}$. The subsequent 
evolution of this state is given by 
%
\begin{align}
\ket*{\Psi_{\vb{r}}(\tau)} &= e^{-i \hat{H}_J \tau} \ket*{h_{\vb{r}}}\nonumber \\
&= e^{-i E_J \tau} \hat{\mathcal{S}} e^{-i\sum_{\vb{k}} \omega_{\vb{k}}^{\mathrm{b}} \tilde{b}_{\vb{k}}^{\dagger} \tilde{b}_{\vb{k}} \tau} \hat{\mathcal{S}}^{-1}\ket*{h_{\vb{r}}}.
\label{Evolution}
\end{align}
Equation \eqref{Evolution} can be solved giving 
\begin{align}
\begin{split}
\ket*{\Psi_{\vb{r}}(\tau)} &= e^{-i E_J \tau} e^{-\sum_{\vb{k}}\abs*{\alpha_{\vb{k}}}^2 \left(1 - e^{-i \omega_{\vb{k}}^{\mathrm{b}} \tau} \right)} \\ & \qquad \times e^{\sum_{\vb{k}}\alpha_{\vb{k}}^* \left(1 - e^{-i \omega_{\vb{k}}^{\mathrm{b}}\tau}  \right) \tilde{b}_{\vb{k}}^{\dagger}}\ket*{h_{\vb{r}}}.
\end{split}
\label{eq:lJwavefunct}
\end{align}
An equivalent wave function was recently obtained in a different context concerning an impurity in a Bose-Einstein condensate~\cite{Shashi2014}.
 Equation \eqref{eq:lJwavefunct} shows that the many-body system is in a coherent state with \emph{time-dependent} coefficients. The spin-wave number statistics is correspondingly time dependent with 
\begin{align}
\begin{split}
\bar{n}_{\vb{k}}(\tau) = 2\abs*{\alpha_{\vb{k}}}^2 [ 1 - \cos(\omega_{\vb{k}}^{\mathrm{b}}\tau) ], \ \sigma_{\vb{k}}(\tau) = \sqrt{\bar{n}_{\vb{k}}(\tau)}.
\end{split}
\label{eq:Timebnumber}
\end{align}
Since the oscillatory terms in Eq.~\eqref{eq:Timebnumber} cancel upon integrating over the Brillouin zone (BZ) in the large time limit $\tau \gg 1/J$, 
the total number of spin-waves $\sum_{\vb{k}} \bar{n}_{\vb{k}}(\tau)$ in the non-equilibrium 
state approaches \emph{twice} the total number in the ground state $\sum_{\vb{k}} \bar{n}_{\vb{k},g}$.

\subsection{Local magnetization}
We now analyse the local magnetization around the hole as a function of the time $\tau$ after it was created.
The magnetization of the TAFM in the absence of the hole is   
\begin{align}
M_{\mathrm{AF}} &= \frac{1}{N} \sum_{\vb{r}} \matrixel*{\mathrm{AF}}{\hat{S}_{\vb{r} }^{z'}}{\mathrm{AF}} 
= \frac{1}{2} -  \frac{1}{N} \sum_{\vb{r}} \matrixel*{\mathrm{AF}}{\hat{s}_{\vb{r}}^{\dagger} \hat{s}_{\vb{r}}}{\mathrm{AF}} \nonumber \\ &= \frac{1}{2} - \frac{1}{N} \sum_{\vb{k}} v_{\vb{k}}^2= 0.2387.
\label{AForder}
\end{align}
As explained in Sec.~\ref{sec:ham}, the  spin operators are rotated locally  and a classical $120^{\circ}$ AFM 
state shown in Fig.~\ref{fig:120deg} would give $M_{\mathrm{AF}} = \frac{1}{2}$. 
Equation \eqref{AForder} shows that quantum fluctuations reduce  this ordering significantly. 

When the hole is created at site $\vb{r}$ and time $\tau=0$, it distorts the magnetization in its surroundings. To quantify this, we calculate the 
local magnetization at site  $\vb{r} + \vb{d}$  
%
\begin{align}
\begin{split}
M(\tau, \vb{d}) = \frac{\expval*{\hat{S}_{\vb{r} + \vb{d}}^{z'}}_{\vb{r}}(\tau)}{M_{\mathrm{AF}}},
\end{split}
\label{eq:Mag}
\end{align}
where $\expval{\hdots}_{\vb{r}}(\tau) = \matrixel*{\Psi_{\vb{r}}(\tau)}{\hdots}{\Psi_{\vb{r}}(\tau)}$ gives the time dependent expectation value. 
With this definition, $M(\tau, \vb{d}) = 1$ if a spin is unaffected by the presence of the hole. 
After Fourier transforming Eq.~\eqref{eq:Mag} to crystal momentum space, we use that since $\ket*{\Psi_{\vb{r}}(\tau)}$ is a coherent state, any spin-wave correlator factors into a product of single operator expectation values. As detailed in App.~\ref{ap:mag}, this gives 
\begin{align}
\begin{split}
M(\tau, \vb{d}) = \frac{M_{\mathrm{AF}} - \abs{\mathcal{A}_{\tau,\vb{d}}\left[u\right] - \mathcal{A}_{\tau,\vb{d}}^*\left[v\right]}^2}{M_{\mathrm{AF}}}
\end{split}
\label{Magtime}
\end{align}
where 
\begin{align}
\begin{split}
\mathcal{A}_{\tau,\vb{d}}\left[u\right] &= \frac{1}{N} \sum_{\vb{q}} e^{i\vb{q} \cdot \vb{d}} \sqrt{N} \expval*{\tilde{b}_{\vb{q}}}_{\vb{r}}(\tau) \  u_{\vb{q}}, \\
\mathcal{A}_{\tau,\vb{d}}\left[v\right] &= \frac{1}{N} \sum_{\vb{q}} e^{i\vb{q} \cdot \vb{d}} \sqrt{N} \expval*{\tilde{b}_{\vb{q}}}_{\vb{r}}(\tau) \  v_{\vb{q}}.
\end{split}
\label{eq:MagAvgs}
\end{align}
and
\begin{align}
\begin{split}
\expval*{\tilde{b}_{\vb{q}}}_{\vb{r}}(\tau) = \alpha_{\vb{k}}^* \left(1 - e^{-i \omega_{\vb{k}}^{\mathrm{b}} \tau} \right).
\end{split}
\label{eq:MagAvgs2}
\end{align}

Equations \eqref{Magtime}-\eqref{eq:MagAvgs2} describe how the magnetic frustration propagates through the lattice after the hole has been created. 
For long times $\tau\gg 1/J$, the oscillatory terms in Eq.~\eqref{eq:MagAvgs} cancel upon integration over the BZ so  that $M(\tau, \vb{d})$
  tends to an asymptotic value. 
Since the magnetization only depends on the state through $\expval{b_{\vb{q}}}_{\vb{r}}(\tau)$, this asymptotic magnetization in fact 
coincides exactly with that of the ground state polaron, where $\expval{b_{\vb{k}}}_{\vb{r},g} = \alpha_{\vb{k}}^*$. 
This is somewhat surprising since there are twice as many  spin waves in the lattice for long times as compared to the ground state polaron. Even more explicitly, 
the overlap between the time-dependent state  and the polaron ground state is $|\bra{\Psi_{\vb{r}, g}}\ket*{\Psi_{\vb{r}}(\tau)}|^2=Z = 0.8003 < 1$
since the time evolution just gives a phase factor, which  shows the difference between the two states. 
The fact that the magnetization for long time matches that of the ground state 
is a direct consequence of the factorization of spin-wave correlators in a coherent state, which leads to cancellation of all time-dependent phase factors. 
More generally, the expectation value of any \textit{local} observable, including higher order correlators, will approach that obtained from the ground state in the limit $\tau \gg 1/J$.

\begin{figure}[t]
\centering
\includegraphics[height=0.23\textwidth]{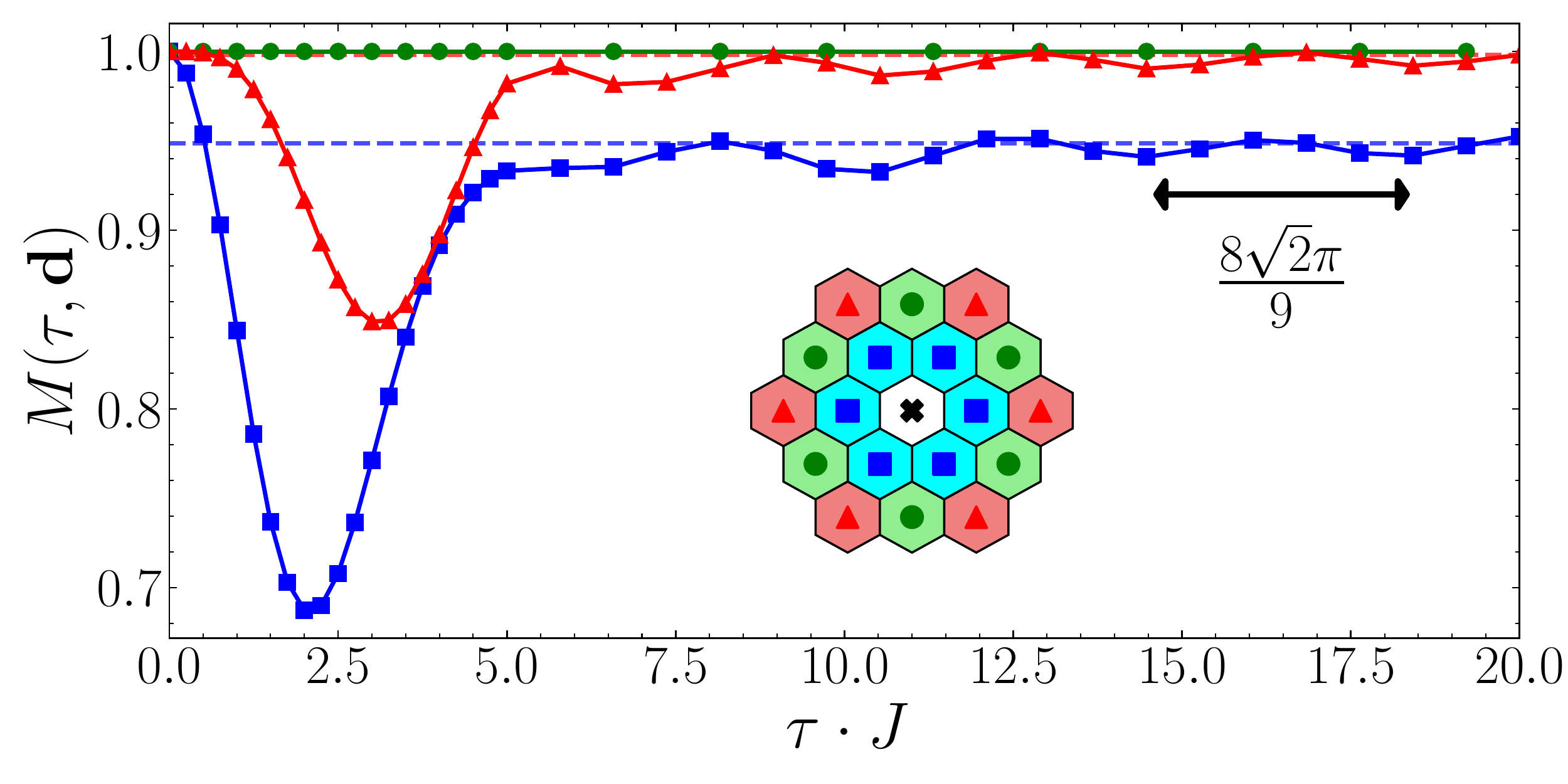}
\caption{Magnetization $M(\tau, \vb{d})$ as a function of time $\tau$ for lattice sites near the hole, indicated by  a black cross. Equivalent lattice sites
 have the same colour. Horizontal  lines show the asymptotic limit obtained from the ground state polaron wave function
 given by Eq.~\eqref{eq:PolaronGroundWaveFunc}. The black horizontal arrow shows a single oscillation period obtained from the stationary phase expansion.}
\label{fig:Mtau}
\end{figure}
\begin{figure*}[t]
\centering
\includegraphics[height=0.26\textwidth]{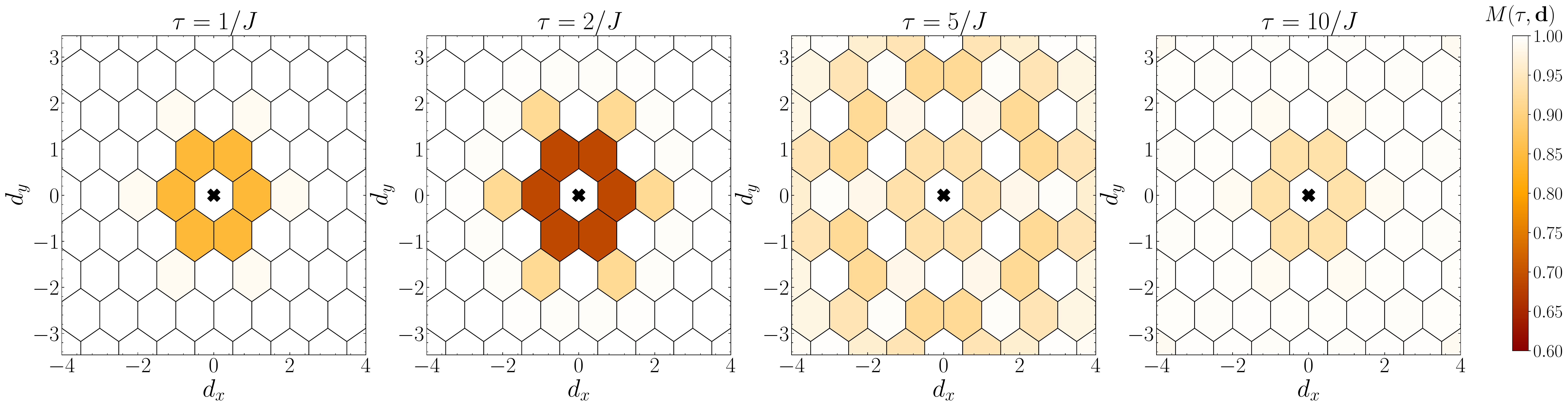}
\caption{Magnetization surrounding a hole (black cross) for four different times after it was created. Each tile represents a spin and its color gives the magnetization. White tiles have an unperturbed magnetization $M(\tau, \vb{d}) = 1$.}
\label{fig:Mlocal}
\end{figure*}
In Fig.~\ref{fig:Mtau},  the magnetization around a stationary hole is plotted as a function of time $\tau$ after it was created. This is obtained 
by  numerically evaluating Eqs.~\eqref{Magtime}-\eqref{eq:MagAvgs2}. 
After the creation of the hole, we see a \kristian{$30 \ \%$} suppression of the magnetization at its nearest neighbours at time $\tau \sim 2/J$. After this initial "burst" of magnetic frustration, the magnetization of the nearest neighbours relax towards a time-independent value coinciding with that of the ground state polaron as discussed above. With some lag, the frustration is then carried through to the  sites further away, which show a similar initial decrease in magnetization although with a smaller amplitude and a subsequent relaxation to the  ground state value. 
 
Since the propagation of the magnetic frustration is carried by spin waves, one  can calculate the period $T_M$  of the oscillations seen in 
  Fig.~\ref{fig:Mtau} using the stationary phase method, which gives $T_M = 2\pi/\omega_{\vb{k}_0}$. Here $\vb{k}_0$ denotes the dominant stationary points where $ \grad \omega_{\vb{k}}^{\mathrm{b}} = 0$, which form an approximate circle around the origin, see Fig.~\ref{fig:disp_bare_0}. From this, we obtain 
\begin{align}
T_M = \frac{8\sqrt{2}\pi}{9J}.
\end{align} 
Figure \ref{fig:Mlocal} shows that this result is indeed confirmed by the numerics. 

Intriguingly,  Fig.~\ref{fig:Mtau} shows that the magnetization at the  green lattice sites is unperturbed by the presence of the hole for all times. 
As detailed in App.~\ref{ap:unperturb}, this  fact follows from the symmetry properties of the triangular lattice. Using those, one can 
show that spins located at horizontal distances  
\begin{align}
\begin{split}
 d_{x} = \frac{3}{2}n \ \forall \ n \in \mathbb{Z}
\end{split}
\label{eq:MCond}
\end{align}
do not feel any disturbing influence from the presence of the hole. For these spins, which all point in the same direction as the spin that was originally 
at the location of the hole, the spin waves destructively interfere so that the magnetization remains unperturbed.


%
%

To further illustrate how the magnetic frustration propagates through the lattice after the hole has been created, we plot 
in Fig.~\ref{fig:Mlocal} the magnetization in a larger region around the hole for four different values of propagation time $\tau$. Here, one clearly observes how a wave of disorder ripples through the lattice. Again, the sites that obey Eq.~\eqref{eq:MCond} are completely unaffected, such that the wave only travels through two of the three sublattices. In the last panel of Fig.~\ref{fig:Mlocal}, the asymptotic steady state is attained, where the spin ordering has relaxed to the polaron ground state value.

\section{SCBA analysis}
\label{sec:SCBA}
Having analysed in detail  the large $J$ regime where hole is stationary, we now examine general values of $J/t$. In this case, the hopping of the hole driven by $\hat H_t$ will lead to additional dressing by magnetic frustration and the problem cannot be solved exactly. 

We analyse the problem via the retarded Green's function of the hole, which in frequency and momentum space reads 
\begin{align}
\begin{split}
G(\vb{k},\omega) = \frac{1}{\omega - \omega_{\vb{k}}^{\mathrm{h}} - \Sigma(\vb{k},\omega)},
\end{split}
\end{align}
where we have suppressed  an infinitesimal positive imaginary part to the frequency $\omega+i\eta^+$. 
The self-energy $\Sigma(\vb{k},\omega)$ is calculated using the self-consistent Born approximation (SCBA), which is known to be very accurate for the equilibrium properties of a hole in a SAFM\cite{Schmitt1988,Kane1989, Martinez1991, Liu1991}. Recently, the accuracy of the SCBA was established for the non-equilibrium case as well where it was shown to agree very well  with experimental data on a SAFM~\cite{nielsen2022}. 
 
\begin{figure}
\centering
\includegraphics[height=0.3\textwidth]{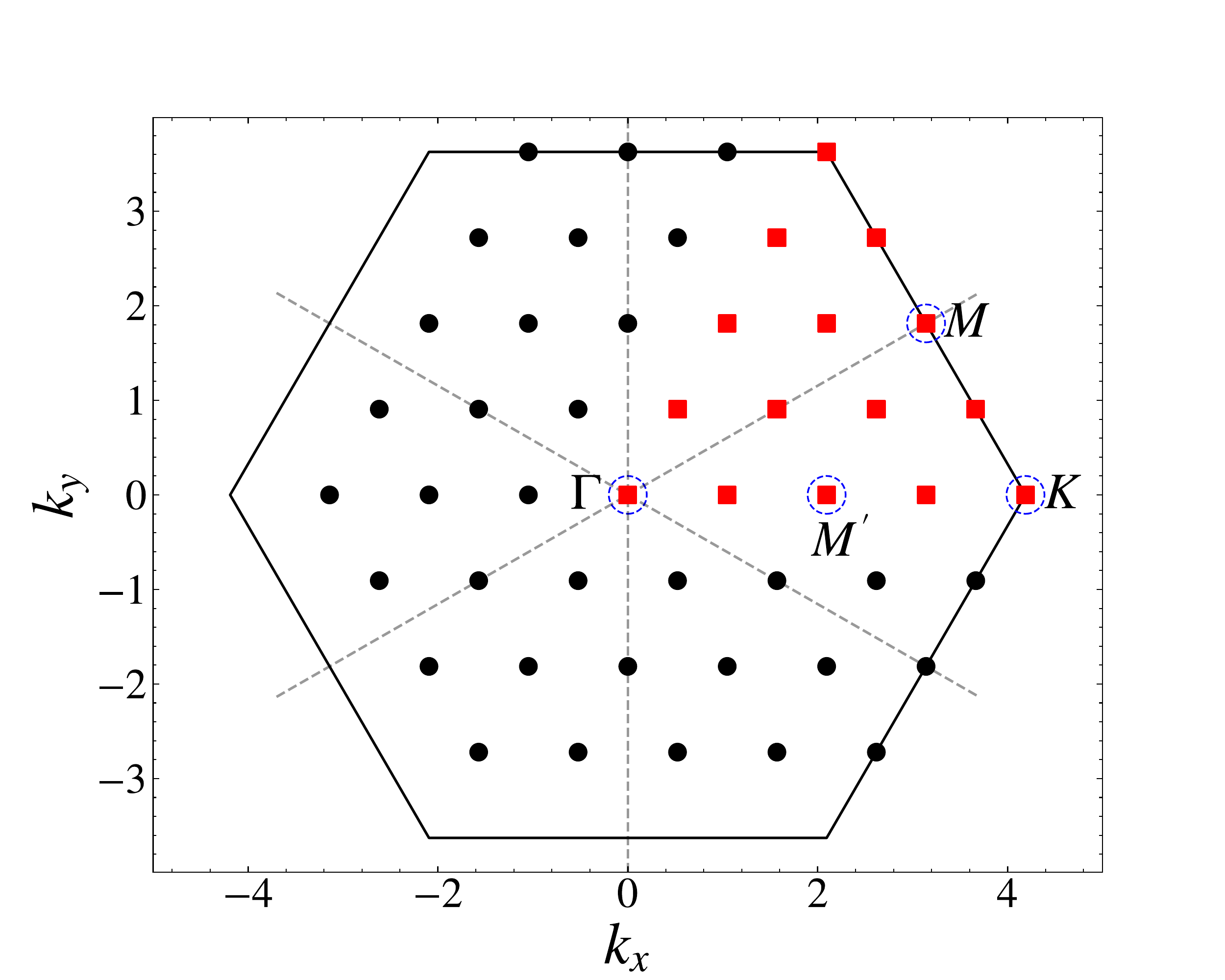}
\caption{Momentum grid in the BZ  for $l = 4$ with the independent states highlighted as red squares.
 The momentum states $\Gamma \equiv (0,0)$, $M' \equiv (2\pi/3, 0)$, $K \equiv (4\pi/3,0)$ and $M \equiv (\pi, \sqrt{\pi/3})$ are shown. The symmetry axes of the structure factors used in deriving 
Eq.~\eqref{eq:MCond} are shown as grey dashed lines.}
\label{fig:lsites}
\end{figure}

In the SCBA, the self-energy is evaluated using all non-crossing diagrams leading to the self-consistent equation
\begin{align}
\Sigma(\vb{k}, \omega) =  \sum_{\vb{q}} \frac{\abs{V_{\vb{k},\vb{q}}}^2}{\omega - \omega_{\vb{q}}^{\mathrm{b}} - \omega_{\vb{k}+\vb{q}}^{\mathrm{h}} - \Sigma(\vb{k+q}, \omega - \omega_{\vb{q}}^{\mathrm{b}})}.
\label{eq:selfen}
\end{align}
 We solve this on a finite momentum grid by iteration starting from $\Sigma(\vb{k}, \omega) = 0$. The grid is created  assuming a regular hexagonal lattice of side length $l$, with periodic boundary conditions such that the total number of independent lattice sites equals $N = 3 l^2$ \cite{Azzouz1996}. As an example, we show in Fig.~\ref{fig:lsites} the grid for the case $l = 4$.
Due to the $C_3$ rotation and $y$-inversion symmetry of the BZ, we only need to evaluate the slice shown in red in Fig.~\ref{fig:lsites}.
In the numerics,  we use a lattice dimension $l=20$, giving a total of $1200$ spins. 
For convergence,  we use a  broadening $\eta_{+}/\abs{t} = 0.01$.

\begin{figure*}[t!]
\centering
\subfloat[\label{fig:spec_J03_0}]{\raisebox{-33ex}{\includegraphics[width=0.32\textwidth]{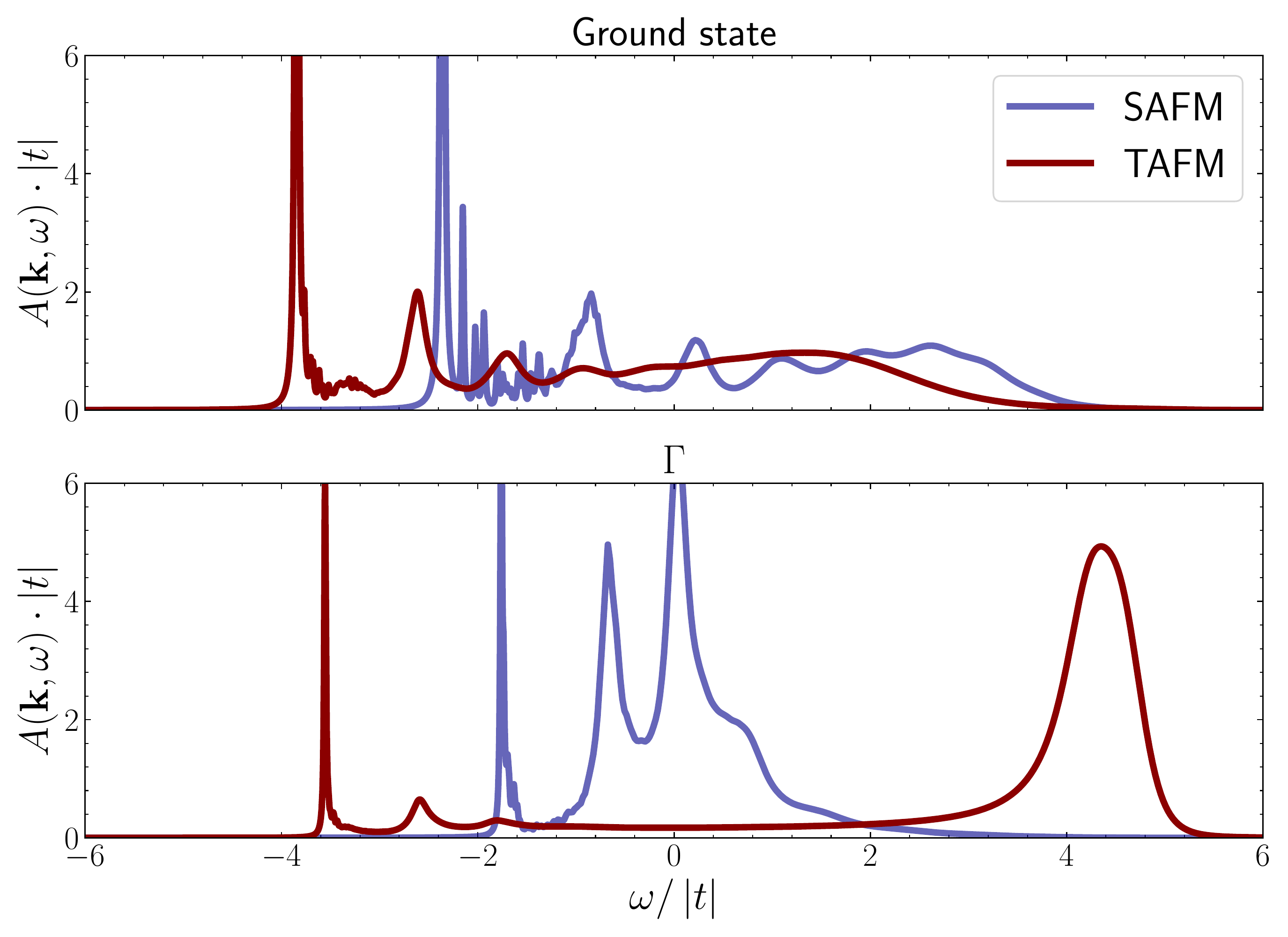}}}
\subfloat[\label{fig:spec_J03_1}]{\includegraphics[width=0.32\textwidth]{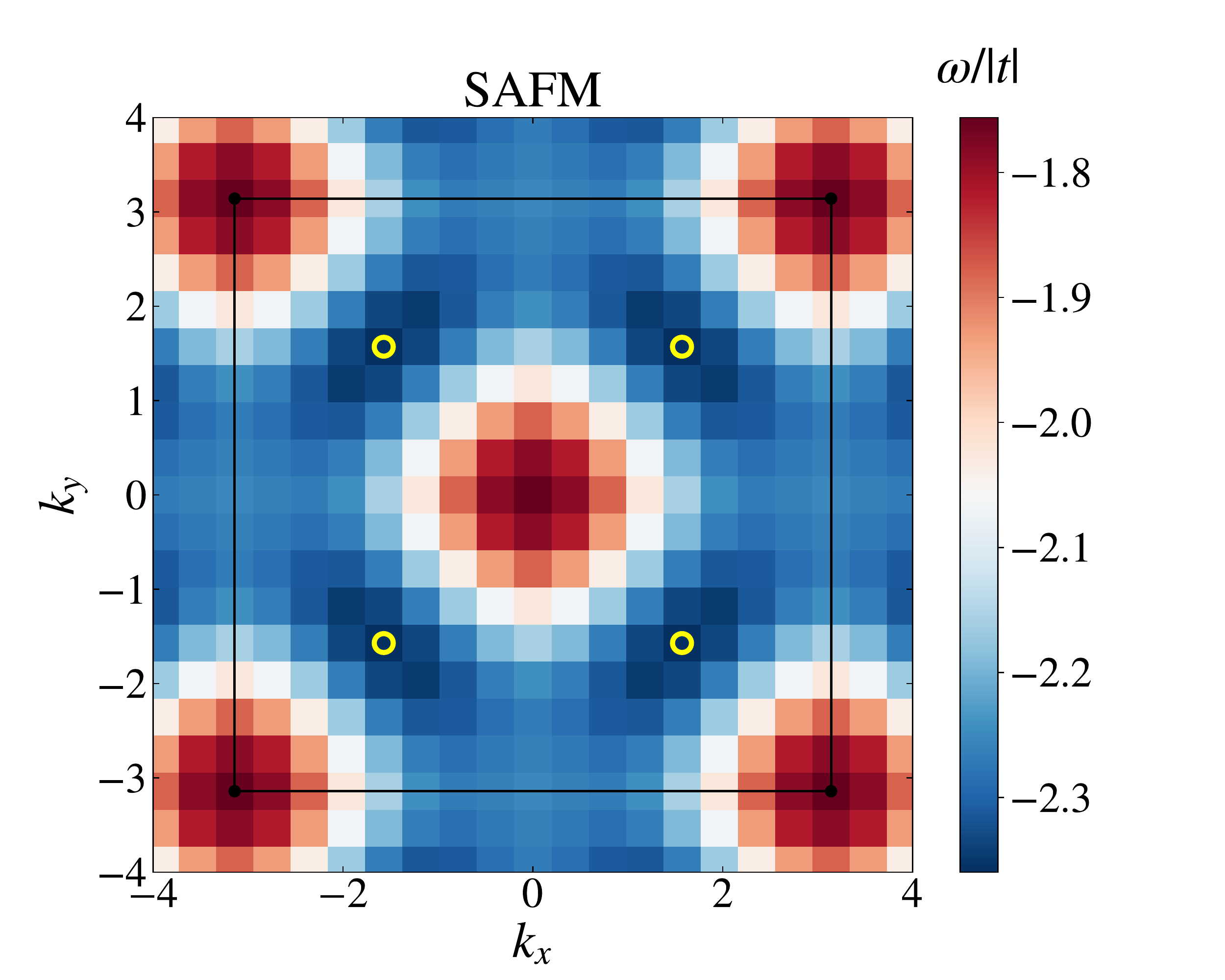}}
\subfloat[\label{fig:spec_J03_2}]{\includegraphics[width=0.32\textwidth]{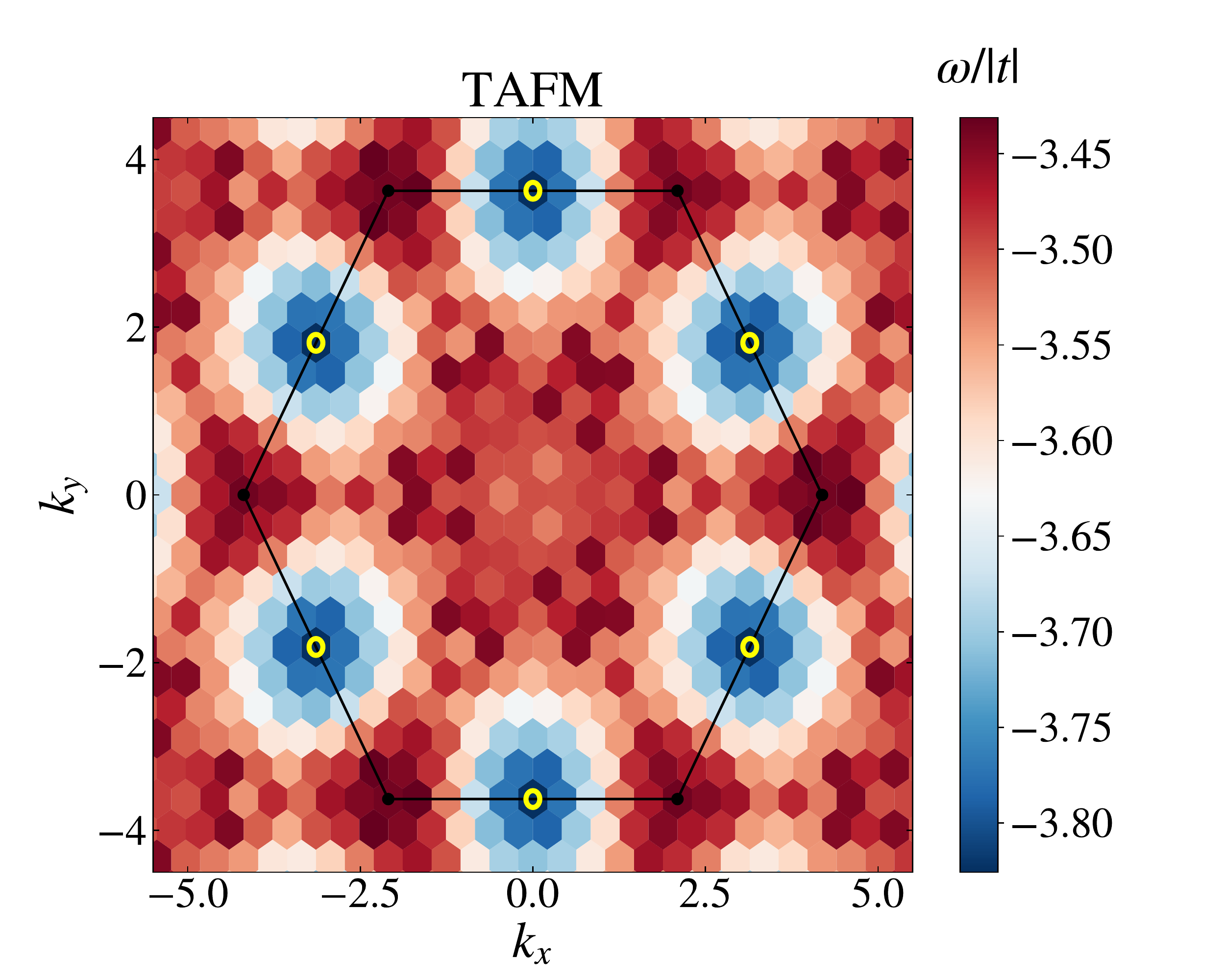}}
\caption{(a) Top panel: The hole spectral functions $A(\vb{k},\omega)$ within SCBA for $J/\abs{t} = 0.3$ and $t<0$ for the ground state, corresponding to the momentum $M$ in the BZ for the TAFM  and the momentum $(\pi/2,\pi/2)$ for the SAFM.  Bottom panel: The spectral functions for zero momentum. 
(b) and (c) show the polaron dispersion in the full BZ for the SAFM.
}
\label{fig:spec_J03}
\end{figure*}

\subsection{Magnetic polarons}
We now explore the properties of the hole  focusing on $t<0$, since the polaron is strongly suppressed except 
for special points in the  BZ  for $t>0$~\cite{Trumper2004}.

In the top panel of Fig. \ref{fig:spec_J03}(a), we plot the hole spectral function $A(\vb{k},\omega)=-2\text{Im}G(\vb{k},\omega)$ for $J/\abs{t} = 0.3$ and the momentum labelled by the point $M$ in the BZ shown in Fig.~\ref{fig:lsites}. We see a clear quasiparticle peak at the 
energy $\omega/|t|\simeq-3.84$, which we interpret as the energy of the magnetic polaron. It turns out that this is the ground state 
 as the polaron energies for other momenta are higher.  For comparison, we also show the spectral function for the hole in a SAFM for the momentum $\vb{k} = (\pi/2, \pi/2)$ exhibiting a polaron peak at the energy $\omega/|t|\simeq-2.4$, which also corresponds to the ground state. Interestingly the ground state spectral functions for the two lattices are quite similar with a clear polaron peak at low energy, followed by a continuum with several smaller resonances.
These peaks have been identified as string excitations in the case of a SAFM, corresponding to the hole oscillating in a linear potential formed by misaligned spins in its wake~\cite{Manousakis2007}, and in analogy we can make the same identification for the case of the TAFM \cite{Trumper2004}.

In the bottom panel of Fig.~\ref{fig:spec_J03}(a), the spectral function is plotted for zero momentum. Again, we see clear polaron 
peaks for both the TAFM and the SAFM. In addition, the spectral function for the TAFM  has a broad new
resonance appearing at $\omega \approx 4.5 \abs{t}$ that  may be interpreted as originating from the bare hole kinetic 
energy, see Fig.~\ref{fig:disp_bare_1}~\cite{Trumper2004}.
If the spectral function is tracked varying the momentum from the origin $\Gamma$ of the BZ towards the point  $M$ on the edge,
we observe that this high energy resonance smoothly evolves into the ground state polaron  consistent with the hole dispersion. 

Finally, we show in Figs.~\ref{fig:spec_J03_1} and \ref{fig:spec_J03_2} the polaron dispersions in the full BZ obtained by solving $\omega_{\vb{k}} - \omega_{\vb{k}}^{\mathrm{h}} - \mathrm{Re} \Sigma(\vb{k}, \omega_{\vb{k}}) = 0$. This illustrates the differences between the energy bands of the polaron in the SAFM and the TAFM.
 
\subsection{Polaron properties as a function of $J/|t|$}
As we discussed above, there is no clear separation between a weak and a strong coupling regime for a hole in a TAFM, 
since the interaction vertex given by Eq.~\eqref{eq:vertex} has a term $\propto J$ in addition to a term $\propto t$. This is in contrast to 
the SAFM, where there is no  interaction term $\propto J$  so that $J/|t|\gg 1$ corresponds to  weak coupling. Therefore, we study in this section the properties of the polaron as a function of $J/|t|$. This will also allow us to make an important benchmarking of the SCBA by comparing it with the exact solution for $J\gg t$ derived in Sec.~\ref{sec:lJ}.
 
\begin{figure}[t!]
\centering
\includegraphics[height=0.4\textwidth]{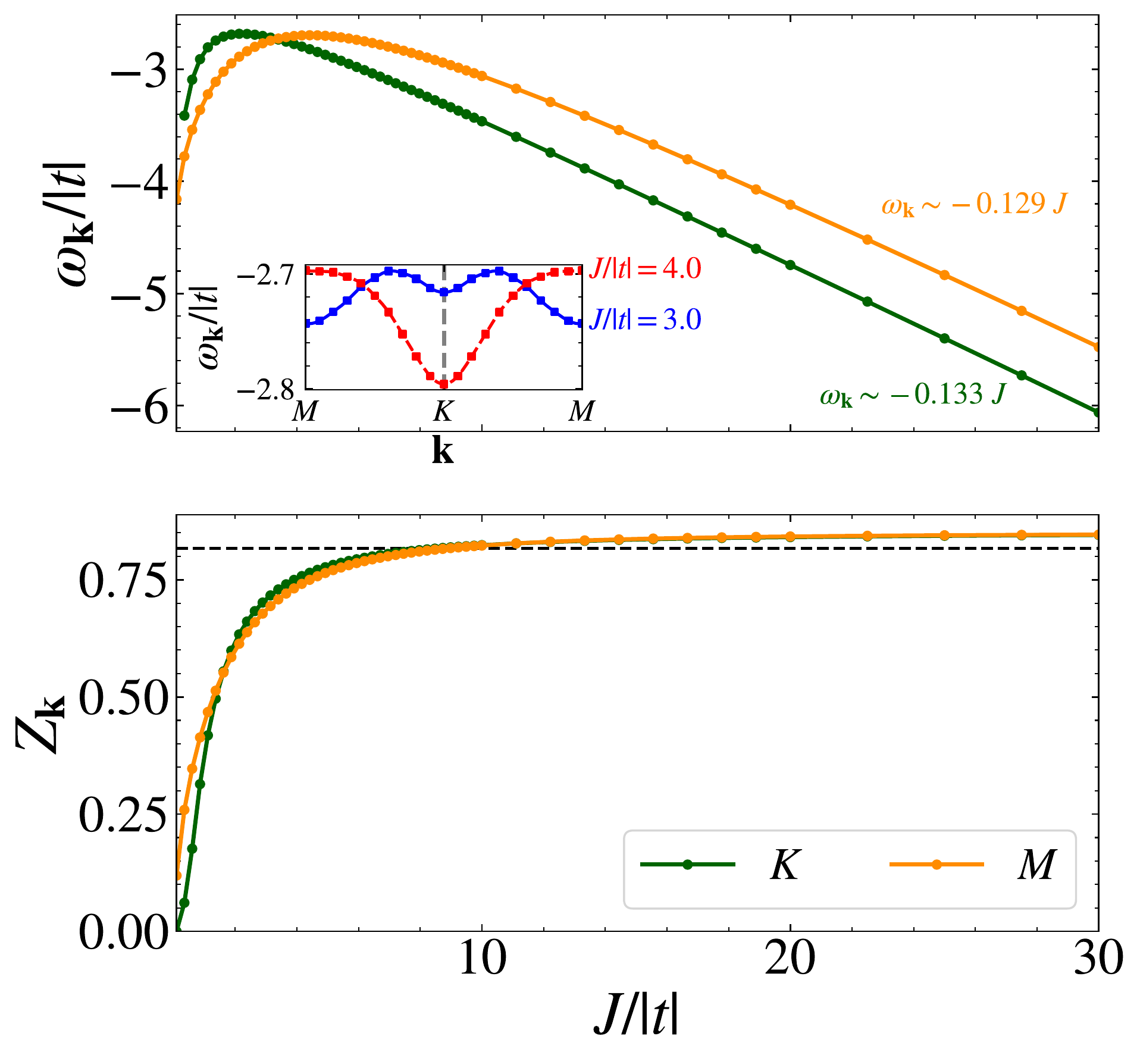}
\caption{The polaron energy $\omega_{\vb{k}}$ (top panel) and quasi-particle residue $Z_{\vb{k}}$ (lower panel) as a function of $J/\abs{t}$ for $t<0$. 
The asymptotic slope of the dispersion for large $J$ is shown for each mode, and the exact large-$J$ quasi-particle residue from the coherent state solution is shown as a dotted line. The inset shows the polaron energy along the edge of the BZ, starting from $M$, passing through $K$, and returning to $M$ for two values of $J/\abs{t}$ on either side of the crossing point of the $M$ and $K$ modes,
 showing the discontinous jump in the ground state momentum.}
\label{fig:scanJ}
\end{figure}

The top panel in Fig.~\ref{fig:scanJ} plots the polaron energy as a function of $J/|t|$  ($t<0$) for the ground state momentum and the momentum 
 at the point $K$ in the BZ, see Fig.~\ref{fig:lsites}. We see that the energies  
 depend non-monotonically on $J/|t|$ exhibiting a maximum for intermediate values. In the limit of large $J/|t|$, the SCBA predicts  the 
 polaron energies approach the asymptotic values $\omega_{\mathbf k}\simeq -0.129J$ and $\omega_{\mathbf k}\simeq -0.133J$ for the $K$ and $M$ points 
  respectively.  Importantly, this  deviates by less than $10\%$ from the exact result given by Eq.~\eqref{GSenergy} for 
  $J/|t|\rightarrow \infty$, which indicates that the SCBA is accurate also for the TAFM in addition to its well-known 
  precision for the SAFM. 
  

 
 In the regime $J/|t|\gg 1$, the momentum dependence of the polaron energy can be calculated perturbatively  to first order in $t$ from the hole kinetic energy term in Eq.~\eqref{eq:Hkspace}. 
 Since the hole dispersion $\omega_\mathbf{k}^\text{h}$ has a minimum at the K point for $t<0$, see Figs.~\ref{fig:disp_bare_1}, the magnetic polaron also has a minimum at this momentum for 
 large $J/|t|$ as can be seen from Fig.~\ref{fig:spec_J03_2}. As a result, the ground state momentum changes discontinuously with increasing $J/|t|$ as is illustrated 
 further in the  inset  of Fig.~\ref{fig:scanJ}.


 The bottom panel in Fig.~\ref{fig:scanJ} plots the  quasiparticle residues of the polarons calculated 
 as $Z_{\vb{k}} = [1 - \partial_\omega \mathrm{Re} \Sigma(\vb{k},\omega)\large|_{\omega = \omega_{\vb{k}}}]^{-1}$~\cite{Mahan2000book}. 
When $J/|t|\rightarrow 0$, the residues go to zero reflecting that a large hopping leads to a strongly dressed hole and an eventual disappearance 
of the magnetic polaron like in the case of the magnetic polaron in a SAFM.  The residues increase with $J/\abs{t}$ and converge to an asymptotic value for $J/\abs{t}\gg 1$, which 
overestimates the exact result $Z =0.8003$ given by Eq.~\eqref{ResidueExact} by just $3\%$. 
In App.~\ref{ap:lim}, we show that this remarkable accuracy of the SCBA in a TAFM can 
can be attributed to the  coherent state coefficients $\alpha_{\vb{k}} = V_{\vb{k}}^J/\omega_{\vb{k}}^{\mathrm{b}}$,
whose absolute value is generally much smaller than unity.


\section{Conclusion and outlook}
\label{sec:conc}
We explored the equilibrium and non-equilibrium dynamics of a hole in a triangular lattice of anti-ferromagnetic spins, using the $t$-$J$ model in  
 a slave fermion representation and linear spin wave theory. 
In the limit of strong magnetic coupling, we derived an exact solution showing that the ground state is a  polaron consisting of a static hole 
 dressed by a coherent state of spin waves, which is  a direct consequence of the geometric frustration of the lattice. We also provided an exact solution for the 
 dynamics ensuing the sudden creation of a hole in a give lattice site. This describes a wave of magnetic frustration emanating from the hole, which leaves one third 
 of the spins in the lattice unperturbed due to a destructive interference between the spin waves. For long times, the magnetization relaxes to that of the ground state 
 polaron. We then benchmarked the SCBA against our exact solution showing that in addition to its well known accuracy on a SAFM, it also gives reliable results on a 
 TAFM. The properties of the magnetic polaron on a TAFM were finally analysed for general coupling strengths and compared  to those on a SAFM. 

Our results motivate several interesting directions for future research. While the linear spin wave theory used here is accurate for 
low energies, the corrections for higher energies are larger in a TAFM than in a SAFM due to geometric frustration~\cite{Chernyshev2009}, and it 
could be interesting to explore their effects on the magnetic polaron. Also, an intriguing but challenging question concerns the properties of a hole in 
a spin liquid phase, which is predicted to exist in a triangular lattice for intermediate values of the on-site repulsion compared to the hopping~\cite{Szasz2020}, 
or in the presence of next-nearest neighbour magnetic coupling~\cite{Drescher2022}. Ultimately, an experimental realization of spin $1/2$ fermions in a triangular lattice
 at low temperatures  using either  atoms in optical lattices~\cite{Yamamoto_2020,Yang2021}, electrons in moir\'e superlattices~\cite{Cao:2018wy,Tang:2020uf,Balents:2020wp,Wu2018}, or Rydberg quantum simulators~\cite{Browaeys:2020tl} will likely 
result in a breakthrough regarding our understanding of  charge motion in geometrically frustrated lattices.

\begin{acknowledgments}
This work has been supported by the Danish National Research Foundation through the Center of Excellence "CCQ" (Grant agreement no.: DNRF156), the Carlsberg Foundation through a Carlsberg Internationalisation Fellowship, and the Dutch Ministry of Economic Affairs and Climate Policy (EZK) as part of the Quantum Delta NL programme.
\end{acknowledgments}

\appendix

\section{Position space Hamiltonian}
\label{ap:Hposfull}

In this appendix we denote all contributions to the second-order $t$-$J$ model Hamiltonian in the slave fermion representation explicitly. First the contribution $\hat{H}_t$ due to the electron hopping is given by,
\begin{align}
\begin{split}
H_t &= \frac{t}{4} \sum_{\vb{i}} \sum_{\vb*{\delta}} \left[ \hat{h}_i \hat{h}_{\vb{i} + \vb*{\delta}}^{\dagger} + \mathrm{H.c.} \right] \\ & + \frac{\sqrt{3}t}{4} \sum_{\vb{i}} \sum_{\vb*{\delta}_+} \big[\hat{h}_i \hat{h}_{\vb{i} + \vb*{\delta}_+}^{\dagger} \hat{s}_{\vb{i} + \vb*{\delta}_+} + \hat{s}_{\vb{i}}^{\dagger} \hat{h}_{\vb{i}} \hat{h}_{\vb{i}- \vb*{\delta}_+}^{\dagger}  \\ & \qquad - \hat{h}_i \hat{h}_{\vb{i} - \vb*{\delta}_+}^{\dagger} \hat{s}_{\vb{i} - \vb*{\delta}_+} -  \hat{s}_{\vb{i}}^{\dagger} \hat{h}_{\vb{i}} \hat{h}_{\vb{i}+ \vb*{\delta}_+}^{\dagger} + \mathrm{H.c.} \big] 
\end{split}
\label{eq:H_t_sf}
\end{align}
The sum over $\vb*{\delta}$ represents a sum over all 6 nearest neighbour vectors. In deriving this Hamiltonian from Eq. \eqref{eq:Ht} we have applied a local Gauge transformation $\hat{c}_{\vb{i}, \sigma}^{\dagger} \rightarrow e^{i\pi} \hat{c}_{\vb{i}, \sigma}^{\dagger}$ for every spin where $i_x = n $ with $n \in \mathbb Z$. This transformation flips the sign on select nearest neighbour vectors, ensuring that the two structure factors $g_{\vb{k}}$ and $\tilde{g}_{\vb{k}}$ are sufficient for expressing the momentum space Hamiltonian. Note that the spin-operators are unaffected by this choice. The expression for $H_J$ is given in Eq. \eqref{eq:HJcubic}, where the quadratic contributions $\hat{H}_s$ are given by,
\begin{align}
\begin{split}
\hat{H}_s &=  \frac{3J}{4} \sum_{\vb{i}} \left[\hat{h}_{\vb{i}}^{\dagger} \hat{h}_{\vb{i}} + 2 \hat{s}_{\vb{i}}^{\dagger} \hat{s}_{\vb{i}}\right] \\ & + \frac{J}{16} \sum_{\vb{i}} \sum_{\vb*{\delta}} \left[\hat{s}_{\vb{i}} \hat{s}_{\vb{i} + \vb*{\delta}}^{\dagger} - 3\hat{s}_{\vb{i}}^{\dagger} \hat{s}_{\vb{i} + \vb*{\delta}}^{\dagger} + \mathrm{H.c.} \right] 
\end{split}
\end{align}

\section{Local magnetization after creation of a hole}
\label{ap:mag}
In this appendix we derive Eq. \eqref{Magtime} from Eq. \eqref{eq:Mag}. First we write the local expectation value of the rotated spin operator into the slave-fermion representation,
\begin{align}
\begin{split}
\expval*{\hat{S}_{\vb{r}+\vb{d}}^{z'}}_{\vb{r}}(\tau) &= \frac{1}{2} - \expval*{\hat{s}_{\vb{r} + \vb{d}}^{\dagger}\hat{s}_{\vb{r} + \vb{d}}}_{\vb{r}}(\tau), \\
&= \frac{1}{2} - \frac{1}{N} \sum_{\vb{k}, \vb{q}} e^{-i(\vb{k} -\vb{q})\cdot(\vb{r}+\vb{d})} \expval*{\hat{s}_{\vb{k}}^{\dagger} \hat{s}_{\vb{q}}}_{\vb{r}}(\tau).
\end{split}
\end{align}
Upon introducing the Bogolubov transformation towards spin-wave operators $\hat{b}_{\vb{k}}$ one can rewrite to obtain,
\begin{align}
\begin{split}
&\expval*{S_{\vb{r}+\vb{d}}^{z'}}_{\vb{r}}(\tau) = \frac{1}{2} - \frac{1}{N} \sum_{\vb{k}} v_{\vb{k}}^2 \\ & - \frac{1}{N} \sum_{\vb{k}, \vb{q}} e^{-i(\vb{k} - \vb{q})\cdot(\vb{r}+\vb{d})} \expval*{\hat{b}_{\vb{k}}^{\dagger} \hat{b}_{\vb{q}}}_{\vb{r}}(\tau) \left( u_{\vb{k}} u_{\vb{q}} + v_{\vb{k}} v_{\vb{q}} \right) \\ & - \frac{1}{N} \sum_{\vb{k}, \vb{q}} e^{-i(\vb{k} - \vb{q})\cdot(\vb{r}+\vb{d})} v_{\vb{k}} u_{\vb{q}} \left(\expval*{\hat{b}_{-\vb{k}} \hat{b}_{\vb{q}}}_{\vb{r}}(\tau) + \expval*{\hat{b}_{\vb{k}}^{\dagger} \hat{b}_{-\vb{q}}^{\dagger}}_{\vb{r}}(\tau) \right) 
\end{split}
\end{align}
One recognizes the expression for $M_{\mathrm{AF}}$ as given in Eq. \eqref{AForder} in the first line. The dependence on $\vb{r}$ disappears after applying the gauge transform $\tilde{b}_{\vb{k}} = \hat{b}_{\vb{k}}e^{i \vb{k} \cdot \vb{r}}$, as expected for a translationally invariant system,
\begin{align}
\begin{split}
&\expval*{S_{\vb{r}+\vb{d}}^{z'}}_{\vb{r}}(\tau) = M_{\mathrm{AF}}  \\ & - \frac{1}{N} \sum_{\vb{k}, \vb{q}} e^{-i(\vb{k} - \vb{q})\cdot\vb{d}} \expval*{\hat{b}_{\vb{k}}^{\dagger} \hat{b}_{\vb{q}}}_{\vb{r}}(\tau) \left( u_{\vb{k}} u_{\vb{q}} + v_{\vb{k}} v_{\vb{q}} \right) \\ & - \frac{1}{N} \sum_{\vb{k}, \vb{q}} e^{-i(\vb{k} - \vb{q})\cdot\vb{d}} v_{\vb{k}} u_{\vb{q}} \left(\expval*{\hat{b}_{-\vb{k}} \hat{b}_{\vb{q}}}_{\vb{r}}(\tau) + \expval*{\hat{b}_{\vb{k}}^{\dagger} \hat{b}_{-\vb{q}}^{\dagger}}_{\vb{r}}(\tau) \right)
\end{split}
\label{eq:MagDeriv1}
\end{align}
Since the coherent state $\ket*{\Psi_{\vb{r}}(\tau)}$ is an eigenstate of $\tilde{b}_{\vb{k}}$, with eigenvalue $\alpha_{\vb{k}}^* \left(1 - e^{-i \omega_{\vb{k}}^{\mathrm{b}} \tau} \right)$, the expectation values in Eq. \eqref{eq:MagDeriv1} factor exactly with respect to the two momenta. Hence the integrals over $\vb{k}$ and $\vb{q}$ are independent and are given by the functions $\mathcal{A}_{\tau,\vb{d}}\left[u\right]$ and $\mathcal{A}_{\tau,\vb{d}}\left[v\right]$ as defined in Eq. \eqref{eq:MagAvgs}

\section{Analysis of unperturbed lattice sites}
\label{ap:unperturb}
As one observes in Figs. \ref{fig:Mtau} and \ref{fig:Mlocal} the presence of the hole has no effect on the local magnetization on one of the three sublattices. This behavior can be derived analytically by noting that the Brillouin zone on the TAFM is exactly mirrored with respect to the two axes $k_x = 0$ and $k_y = \pm k_x /\sqrt{3}$ (see Fig. \ref{fig:lsites}). The structure factors $g_{\vb{k}}$ and $\tilde{g}_{\vb{k}}$ are respectively symmetric and asymmetric with respect to these axes, as can be checked by computing the mirror images,
\begin{align}
\begin{split}
\vb{k}_1 = \begin{pmatrix}
\frac{1}{2}  & \frac{\sqrt{3}}{2} \\
\frac{\sqrt{3}}{2} & - \frac{1}{2}
\end{pmatrix} \vb{k}, \qquad \vb{k}_2 = \begin{pmatrix}
-1  & 0 \\
0 & 1
\end{pmatrix} \vb{k},
\end{split}
\end{align}
which gives $g_{\vb{k}} = g_{\vb{k}_1} = g_{\vb{k}_2}$ and $\tilde{g}_{\vb{k}} = -\tilde{g}_{\vb{k}_1} = -\tilde{g}_{\vb{k}_2}$. Consider now the integrals $\mathcal{A}_{\tau,\vb{d}}\left[u\right]$ and $\mathcal{A}_{\tau,\vb{d}}\left[v\right]$ which set the $\vb{d}$ dependence of the magnetization. Excluding the $\vb{d}$ dependent phase factor, one can use the stated properties of the structure factors to show that the integrand is always asymmetric with respect to the symmetry axes. Hence, for $\abs{\vb{d}} = 0$, the total integral should always vanish exactly. For finite $\abs{\vb{d}}$, the integral will vanish if $e^{i \vb{k} \cdot \vb{d}}$ is symmetric with respect to any of the symmetry axes, i.e. if $e^{i \vb{k} \cdot \vb{d}} = e^{i \vb{k}_1 \cdot \vb{d}}$ or $e^{i \vb{k} \cdot \vb{d}} = e^{i \vb{k}_2 \cdot \vb{d}}$. Solving for $\vb{d}$ will give Eq. \eqref{eq:MCond}, which correspondingly captures all the sites where the symmetry of the lattice dictates that no disturbing influence of the hole will be felt. Note that in the $120^{\circ}$ Néel AF state the sites where \eqref{eq:MCond} holds correspond with the sites where the spins are aligned with the spin that was originally at the site of the hole, i.e. the sites where $\theta = \vb{Q} \cdot \vb{d} = 0$.

\section{SCBA results for positive $t$}

\begin{figure}
\centering
\includegraphics[height=0.4\textwidth]{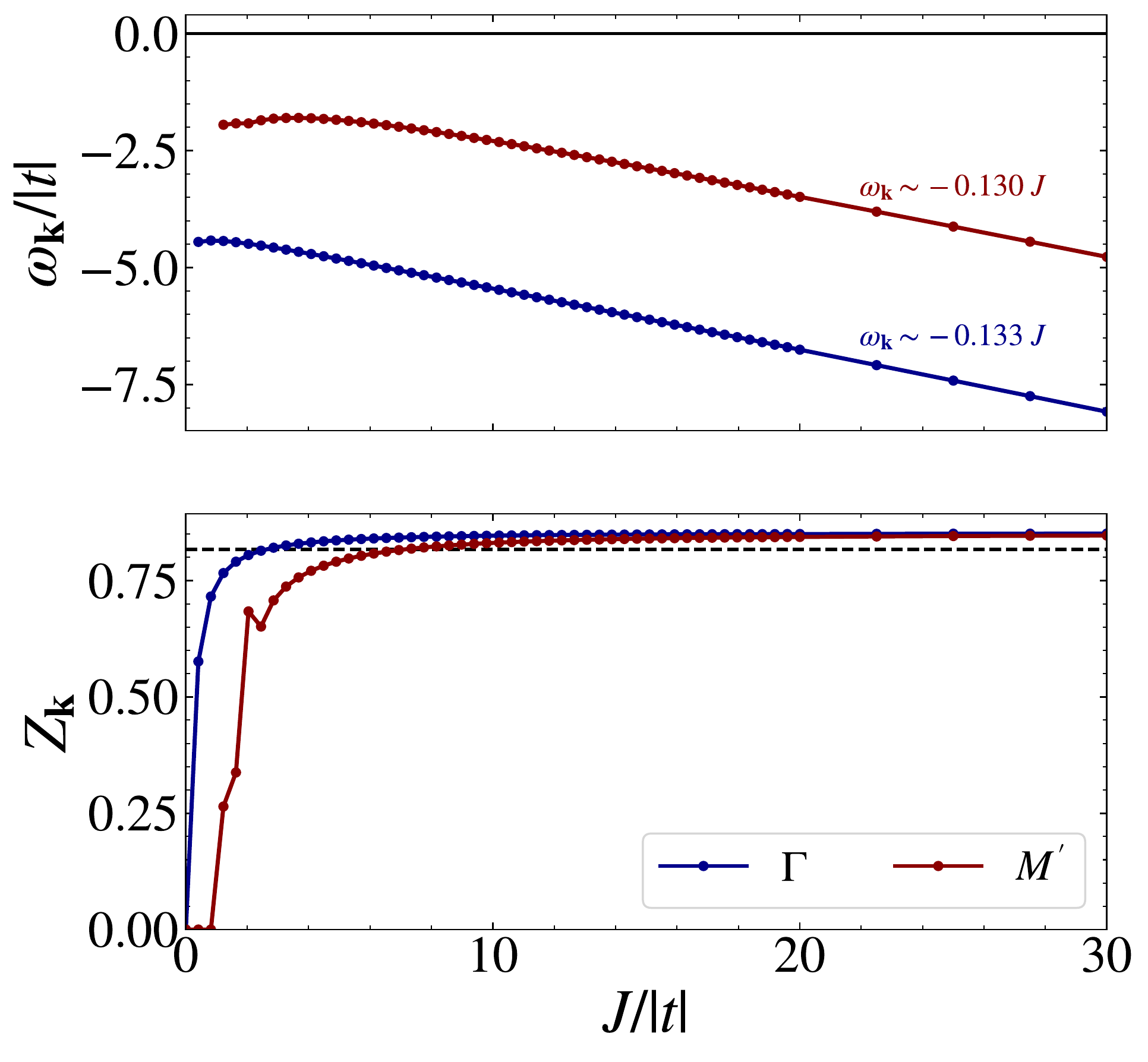}
\caption{The polaron energy $\omega_{\vb{k}}$ (top panel) and quasi-particle residue $Z_{\vb{k}}$ (lower panel) as a function of $J/\abs{t}$ for $t>0$.}
\label{fig:tplus}
\end{figure}

In this section we give some additional SCBA results for the case $t>0$. They are collected in Fig. \ref{fig:tplus} which can be directly compared with Fig. \ref{fig:scanJ}. As expected, the results for large $J/t$ converge to the coherent state results of Sec. \ref{sec:lJ} with some small error due to the SCBA approximation. In contrast with the case $t<0$, the polaron ground state is situated at the zero momentum $\Gamma$ regardless of the value of $J$, which matches the ground state of the uncoupled spin-waves and holes (see Fig. \ref{fig:disp_bare}). Consequently the $t>0$ magnetic polaron lacks a ground state crossing as was observed for negative $t$.
In the small to intermediate $J/\abs{t}$ regime the excited $M'$ polaron state abruply disappears when $J/t\lesssim 1$. This sudden suppression of the quasi-particle weight has been analysed in detail in Ref. \cite{Trumper2004}, using a similar approach but neglecting the static $J$-dependent spin-wave interaction. Our model reproduces their findings, altered by a small inflection point in the quasi-particle weight around the critical value.
\section{$J$ scaling of excited polaron modes}
\label{ap:othersol}
\begin{figure*}[t]
     \centering
     \subfloat[\label{fig:dispt-1J}]{\includegraphics[width=0.32\textwidth]{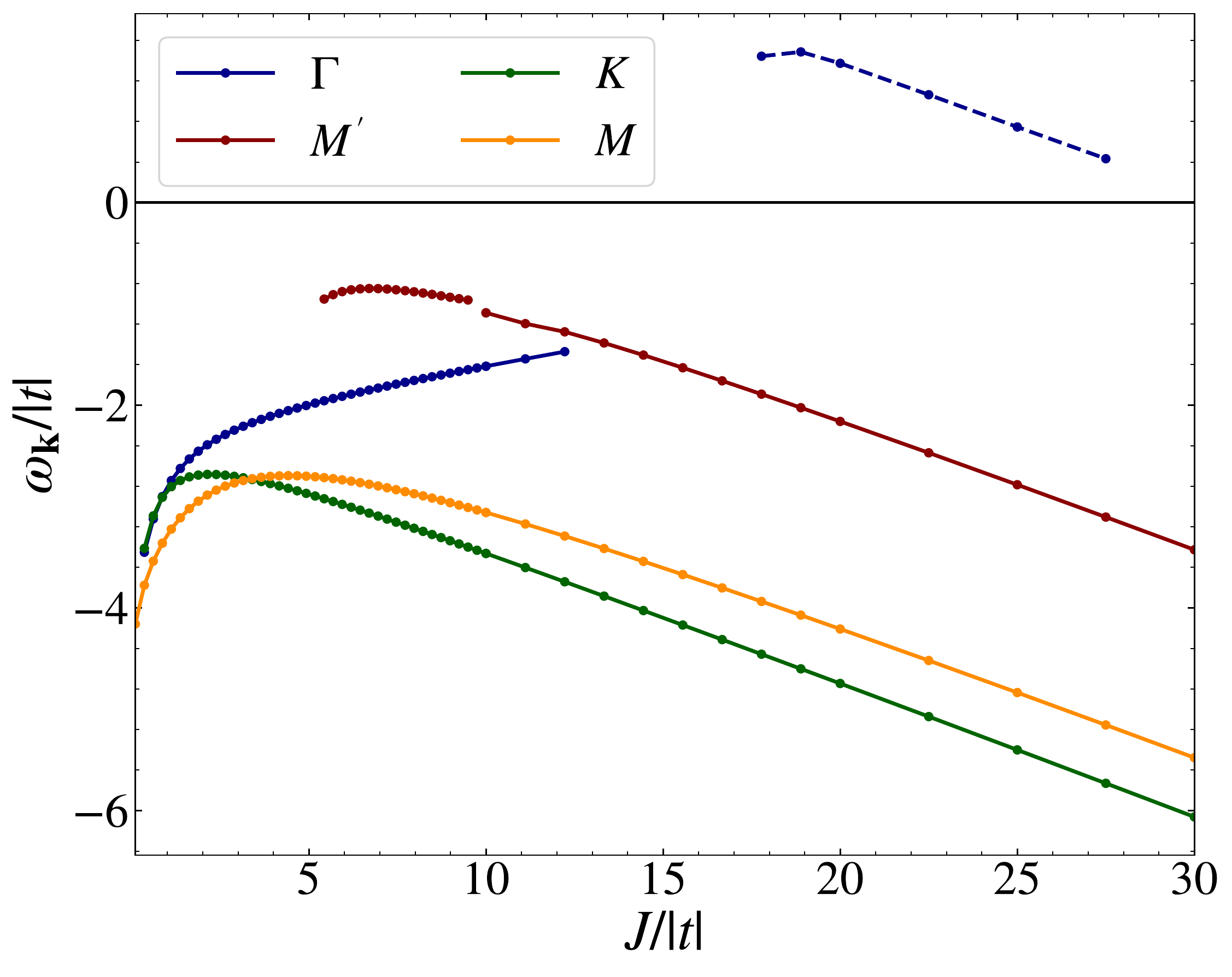}}
     \subfloat[\label{fig:QPwt-1J}]{\includegraphics[width=0.32\textwidth]{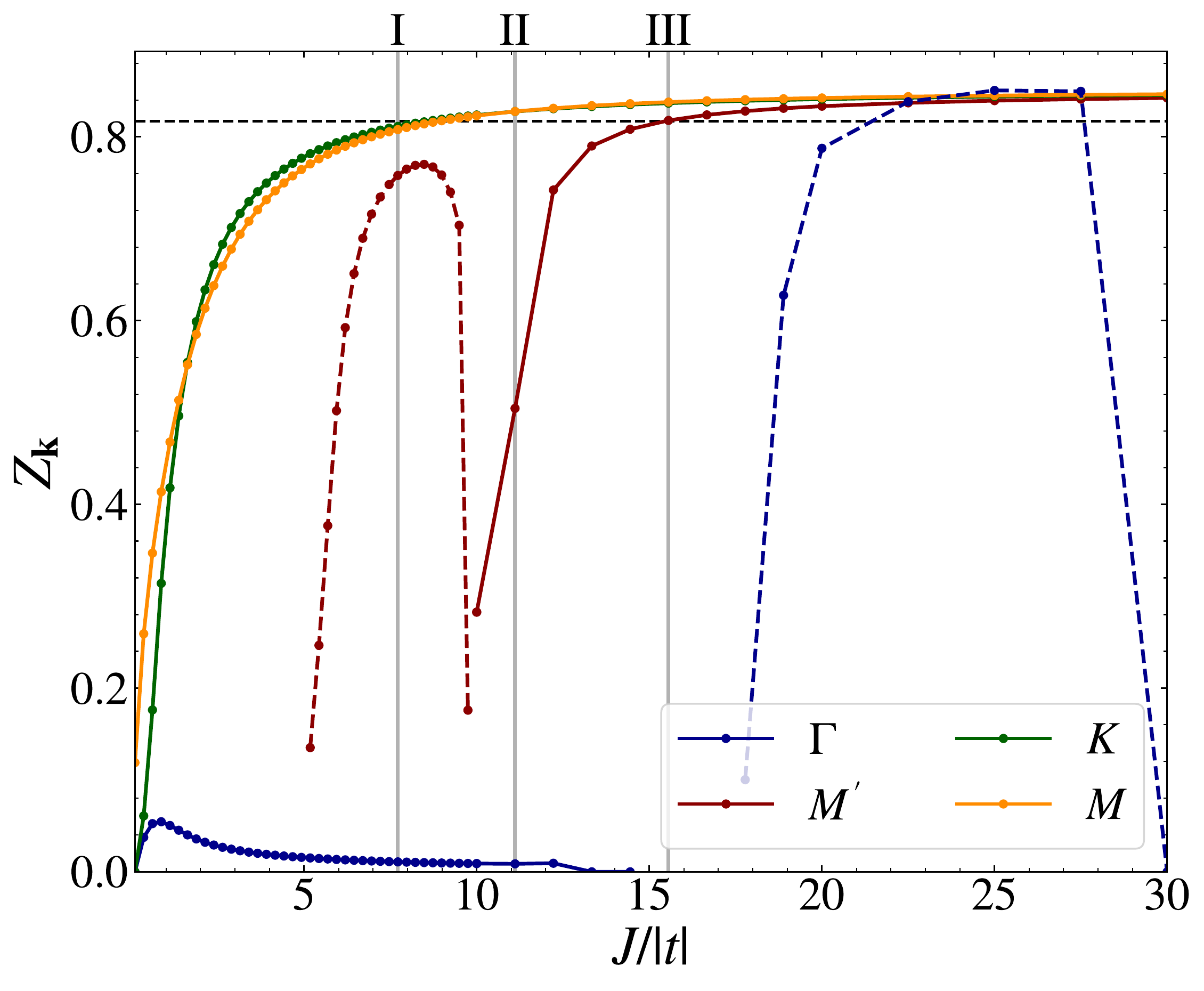}}
     \subfloat[\label{fig:Mp_comp_t-1}]{\includegraphics[width=0.32\textwidth]{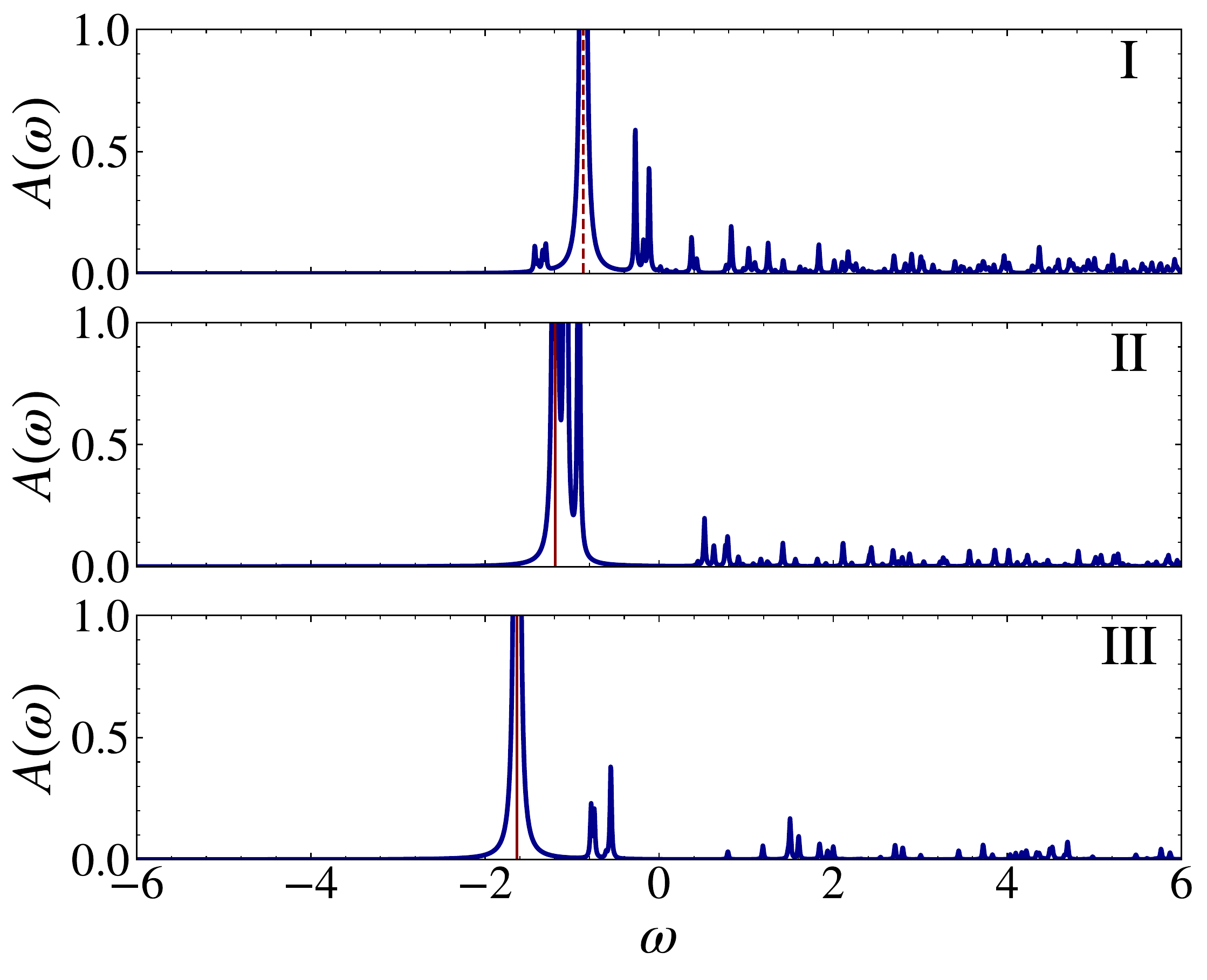}}
     \caption{Behavior of the magnetic polaron in all four modes highlighted in Fig. \ref{fig:lsites} as a function of $J/\abs{t}$, for the case $t < 0$. In Figs. (a) and (b) the polaron energy and spectral weight are shown, where each mode is distinguished with a unique colour. In Fig. (c) we show the full spectral function at the $M'$ mode for three distinct values of $J$ encoded as points I, II and III. These are shown in Fig. (b) with vertical grey lines.}
\label{fig:t-1J}
\end{figure*}

In this appendix we present some additional results for the scaling of the polaron modes with $J$, specifically considering the case $t<0$ and the scaling of the modes $\Gamma$ and $M'$ in the inner parts of the BZ. Scans of the polaron energy and spectral weight similar to Fig. \ref{fig:scanJ} in the main text are shown in Figs. \ref{fig:dispt-1J} and \ref{fig:QPwt-1J}.

Consider first the $M'$ mode. As briefly mentioned in the main text, a robust quasi-particle solution already appears for intermediate value of $J$, although it does not develop smoothly into the coherent state. First around $J/\abs{t} \approx 10$ we observe a sharp feature in which the spectral weight of the polaron is strongly supressed. To understand this better we plot in Fig. \ref{fig:Mp_comp_t-1} the full spectral function $A(\vb{\omega})$ at the $M'$ mode for three different values of $J$ around the suppressive feature, denoted with I, II and III and also shown in Fig. \ref{fig:QPwt-1J}. For $J/\abs{t} < 10$ there exists a clear polaron solution as shown by the red dotted line. One will note that just below this peak there exist a couple of smaller resonances where $\abs{\omega_{\vb{k}} - \omega_{\vb{k}}^{\mathrm{h}} - \mathrm{Re} \Sigma(\vb{k}, \omega_{\vb{k}})}$ has local minima. As $J$ is increased these minima move closer to zero, crossing through zero at some critical value of $J$ leading to new quasi-particle solutions just below the solution we have highlighted. At point II these different closely spaced solutions lead to a break-up of the quasi-particle peak, creating a strong suppression of spectral weight as observed in Fig. \ref{fig:QPwt-1J}. As $J$ is increased further, a new single peak emerges that will subsequently develop into the coherent state, as shown in the spectral function at point III. The $\Gamma$ mode also shows different non-trivial behavior. For small and intermediate values of $J$ there exists a quasi-particle solution with low spectral weight, also observed in Ref. \cite{Trumper2004}. As $J$ is increased this solution moves to higher energy, suggesting that it develops into some excited polaron state. Then at some point it is damped completely in favour of a new solution that quickly grows in spectral weight towards the asymptotic coherent state value. This solution however follows a similar curve as observed for the $M'$ mode, with a strong suppression of spectral weight around $J/\abs{t} \approx 30$ before developing into the coherent state solution outside of the range we plot. Evidently the behavior of excited polaron modes in our model is quite complicated, and we note that it may be influenced by finite size effects that persist on our 1200 spin lattice. 

\section{Comparison SCBA with large $J$ limit}
\label{ap:lim}

In this appendix we show that the coherent state self-energy equation reproduces the SCBA equation up to first order in the dimensionless coefficient $\abs{\alpha_{\vb{k}}} = \abs*{V_{\vb{k}}^J/\omega_{\vb{k}}^{\mathrm{b}}}$. First we can use Eq. \eqref{eq:S} to obtain the time dependent hole  Green's function $G(\tau) = -i \theta(\tau) \matrixel*{h}{e^{-i H \tau}}{h}$. Using the same techniques as applied in deriving Eq. \eqref{eq:lJwavefunct} we find,
\begin{align}
\begin{split}
G(\tau) = -i \theta(\tau) e^{-i E_J \tau} e^{-\sum_{\vb{k}}\abs*{\alpha_{\vb{k}}}^2 \left(1 - e^{-i \omega_{\vb{k}}^{\mathrm{b}} \tau} \right)}.
\end{split}
\end{align}
Now take the derivative with respect to $\tau$ on both sides and Fourier transform to frequency space. Then we obtain the following self-consistent equation for the Green's function,
\begin{align}
\begin{split}
G^{-1}(\omega) = \frac{\omega - E_J}{1 + \sum_{\vb{k}} \frac{\abs*{V_{\vb{k}}^J}^2}{ \omega_{\vb{k}}^{\mathrm{b}}} G_{\vb{r}}(\omega - \omega_{\vb{k}}^{\mathrm{b}})}. 
\end{split}
\end{align}
As expected the Green's function diverges at the polaron ground state energy $\omega = E_J$, and is momentum independent. From the Green's function we obtain a self-consistent equation for the self-energy,
\begin{align}
\begin{split}
\Sigma_{\vb{r}}(\omega) = \frac{\omega\sum_{\vb{k}} \frac{\abs{V_{\vb{k}}^J}^2}{\omega_{\vb{k}}^{\mathrm{b}}} \frac{1}{\omega - \omega_{\vb{k}}^{\mathrm{b}} - \Sigma(\omega - \omega_{\vb{k}}^{\mathrm{b}})} + E_J}{1 + \sum_{\vb{k}} \frac{\abs{V_{\vb{k}}^J}^2}{\omega_{\vb{k}}^{\mathrm{b}}} \frac{1}{\omega - \omega_{\vb{k}}^{\mathrm{b}} - \Sigma(\omega - \omega_{\vb{k}}^{\mathrm{b}})}}.
\end{split}
\end{align}
Note that for $V_{\vb{k}}^J = 0$ this equation reduces trivially to the associated SCBA equation. A Taylor expansion to first order in $\alpha_{\vb{k}}$ gives, 
\begin{align}
\begin{split}
\Sigma_{\vb{r}}(\omega) \approx \omega\sum_{\vb{k}} \frac{\abs{V_{\vb{k}}^J}^2}{\omega_{\vb{k}}^{\mathrm{b}}} \frac{1}{\omega - \omega_{\vb{k}}^{\mathrm{b}} - \Sigma(\omega - \omega_{\vb{k}}^{\mathrm{b}})} + E_J
\end{split}
\end{align}
Or, upon substituting $E_J$ and rewriting,
\begin{align}
\begin{split}
\Sigma_{\vb{r}}(\omega) &\approx \sum_{\vb{k}} \frac{\abs{V_{\vb{k}}^J}^2}{\omega - \omega_{\vb{k}}^{\mathrm{b}} - \Sigma(\omega - \omega_{\vb{k}}^{\mathrm{b}})} \\& \quad+ \sum_{\vb{k}} \frac{\abs{V_{\vb{k}}^J}^2}{\omega_{\vb{k}}^{\mathrm{b}}} \frac{\Sigma(\omega - \omega_{\vb{k}}^{\mathrm{b}})}{\omega - \omega_{\vb{k}}^{\mathrm{b}} - \Sigma(\omega - \omega_{\vb{k}}^{\mathrm{b}})}.
\end{split}
\end{align}
The first line is just the SCBA equation \eqref{eq:selfen} with $t = 0$. The second line encodes higher order corrections due to crossing diagrams neglected in the SCBA approach, which scale with higher powers of $\abs*{\alpha_{\vb{k}}}$. 

\bibliography{References}

\end{document}